\documentclass[%
reprint,longbibliography, %To show titles in the bibiliography.
superscriptaddress,
showpacs,%preprintnumbers,
 amsmath,amssymb,
 pra,
]{revtex4-1}

\usepackage[utf8]{inputenc}
\usepackage[english]{babel}

\usepackage{braket}
\usepackage{graphicx,color,tabularx}
\usepackage[colorlinks=true,citecolor=blue,linkcolor=blue,urlcolor=blue]{hyperref}

\newcommand{\unit}[1]{\ensuremath{\, \mathrm{#1}}}

\DeclareMathOperator{\sinc}{sinc}

\newcommand{\corr}[1]{\textcolor{black}{#1}}
\def\e{\hbox{e}}

\begin{document}

\title{Asymmetric Talbot-Lau interferometry for inertial sensing}

\author{Simone Sala}
\affiliation{Dipartimento di Fisica, Università degli Studi di Milano, I-20133 Milano, Italy}
\affiliation{INFN Sezione di Milano, I-20133 Milano, Italy}

\author{Marco Giammarchi}
%\email{marco.giammarchi@mi.infn.it}
\affiliation{INFN Sezione di Milano, I-20133 Milano, Italy}
\affiliation{Albert Einstein Center for Fundamental Physics Laboratory for High Energy Physics University of Bern Sidlerstrasse, 5 CH-3012 Bern}

\author{Stefano Olivares}
\email{stefano.olivares@fisica.unimi.it}
\affiliation{Dipartimento di Fisica, Università degli Studi di Milano, I-20133 Milano, Italy}

\begin{abstract}
We study in detail a peculiar configuration of the Talbot-Lau matter wave interferometer, characterised by unequal distances between the two diffraction gratings and the observation plane. We refer to this apparatus as the ``asymmetric Talbot-Lau setup''. Particular attention is given to its capabilities as an inertial sensor for particle and atomic beams, also in comparison with the classical moiré deflectometer. The present analysis is motivated by possible experimental applications in the context of antimatter wave interferometry, including the measurement of the gravitational acceleration of antimatter particles. To support our findings, we have performed numerical simulations of realistic particle beams with varying speed distributions. 
\end{abstract}


\maketitle

\section{Introduction}

Inertial sensors for particle beams based on material transmission gratings exist and have been studied extensively (see for example \cite{Batelaan199785,atomintgeneric}). Commonly, these devices are moiré deflectometers \cite{moiree}: two-grating setups operating in the classical regime, with the particles following ballistic trajectories and producing geometrical shadow fringe patterns. The presence of a constant and uniform force in the transverse direction (corresponding to an acceleration $a$) induces a displacement $\Delta x  \propto a T_1 ^2$ in the fringe pattern, where $T_1$ is the time of flight between the two gratings. It is known that the Talbot-Lau matter-wave interferometer \cite{clauser} also possesses the same inertial sensitivity \cite{arndtonatomint}. Unlike the moiré deflectometer, this device operates in the quantum diffraction regime; therefore the properties of the interference pattern depend on the de Broglie wavelength of the interfering particles \cite{amattint}. 

In this paper we aim to investigate the precise behaviour of the displacement $\Delta x$ as a function of $a$ and $T_1$, in both the classical and quantum regime. More in details, we use the Wigner function formalism \cite{arndtonatomint, LeeWigner} to study the statistical interference pattern produced by a general Talbot-Lau interferometer in the presence of an external force. First we recover the result that period-magnifying interferometers can be realised under the appropriate resonance conditions \cite{patorski, concepts}, and perform a systematic analysis of their features (see section \ref{sec:general_description}).

We then proceed to investigate the inertial sensitivity properties of these peculiar setups, which will be referred to as \emph{asymmetric} Talbot-Lau interferometers. Our analysis of the inertial sensitivity shows that the absolute fringe displacement scales quadratically with the magnification factor in asymmetric configurations. This is discussed in section \ref{sec:inertial_sensitivity}.

In a previous paper we discussed the possible experimental applications of Talbot-Lau setups in the interferometry of antimatter particles (e.g., positrons and positronium (Ps) atoms \cite{amattint}). An interesting development can be the measurement of the gravitational acceleration of neutral antimatter with Talbot-Lau interferometers, as opposed to classical moiré deflectometers \cite{aegisproposal}. With this application in mind, we compare the standard and asymmetric Talbot-Lau interferometers of the same total length in order to establish whether there is a systematic advantage in using an asymmetric setup, with respect to period magnification and inertial sensitivity (see sections \ref{sec:asymmetric_setups} and \ref{sec:inertial_sensitivity}). 
In this case we find that the gain in the absolute inertial fringe displacement is effectively limited, and also that the relative displacement $\Delta x / d_3$ (where $d_3$ is the fringe period) vanishes for high magnification factor . The magnitude of the fringe period instead scales favourably with the magnification factor, which can be a relevant advantage when the experimental resolution is a concern, as it is for example the case of low energy electrons or positrons \cite{amattint}.

For these reasons, the best trade-off when designing an inertial sensor based on a Talbot-Lau interferometer has to be found given the specific properties of the particle beam. The numerical analysis of section \ref{sec:Monte_Carlo} indicates that asymmetric configurations are useful for this purpose due to their peculiar properties.

\section{General description of a Talbot-Lau interferometer} \label{sec:general_description}

A satisfactory theoretical treatment of grating matter-wave interferometers exploits the analogy with classical scalar diffraction theory \cite{amattint, patorski}. This is justified by the formal correspondence existing between the time evolution of the wave function calculated via the Schr\"{o}dinger equation, and the Fresnel-Kirchhoff diffraction integral \cite{wolf2,patorski} for a classical scalar field of wavelength $\lambda = \lambda(v) = h/(mv)$, where $m$ and $v$ are the mass and velocity of the particle, respectively.  

\begin{figure}[htb]
\includegraphics[width=0.4\textwidth]{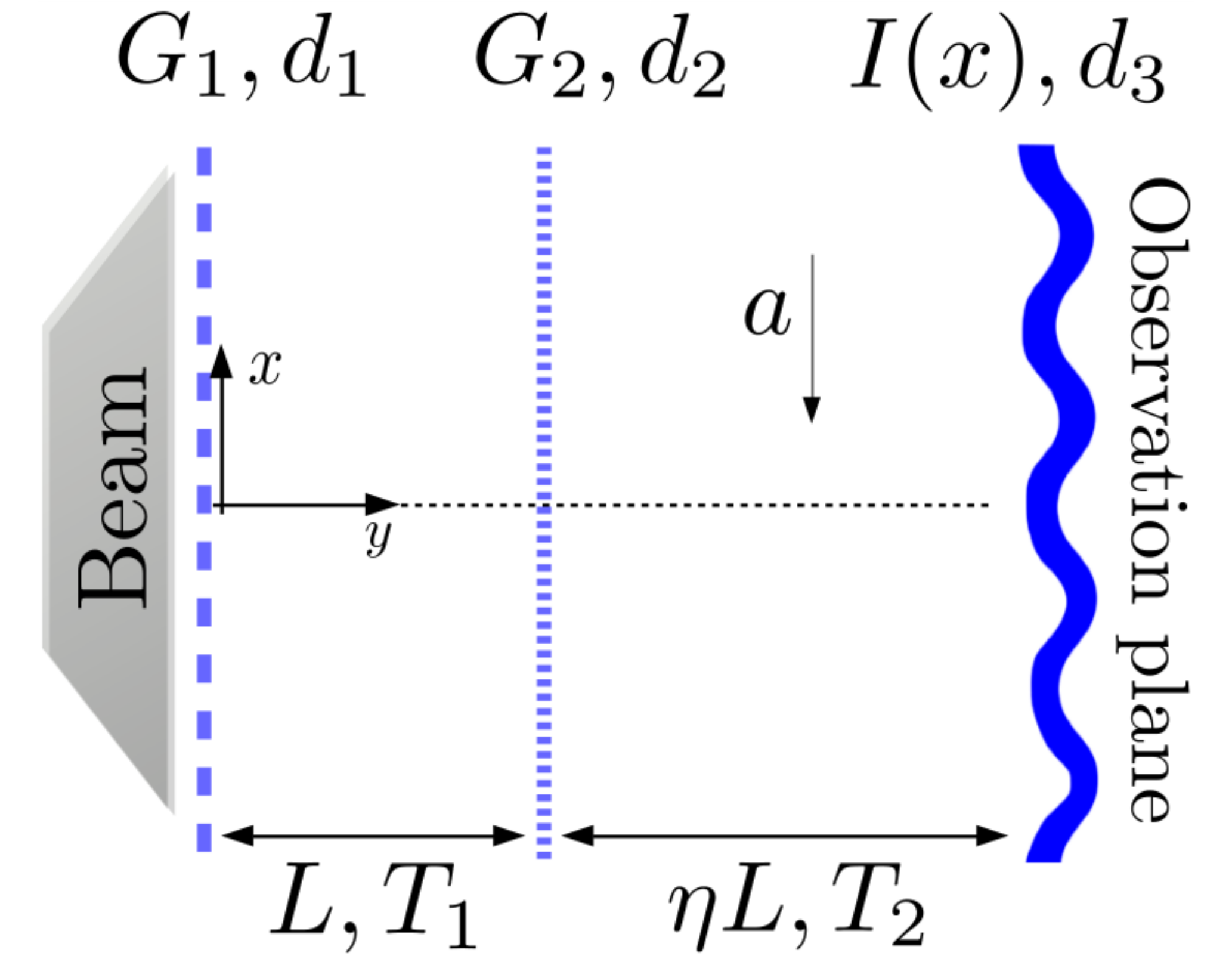}
\vspace{-0.3cm}
\caption{\corr{General Talbot-Lau setup in the presence of an external acceleration $a$ acting on the particles along the $x$-direction. The particles travel along the $y$-axis with longitudinal speed $v$. Two diffraction gratings $G_1$  and $G_2$ of period $d_1$ and $d_2$ are located on the $y=0$ and $y=L$ planes, respectively. The detection plane is placed at $y= (1+\eta) L$, where an interference fringe pattern $I(x)$ with period $d_3$ is formed. The time of flight between the two gratings is $T_1=L/v$ while $T_2 = \eta T_1$ is the time of flight between $G_2$ and the detection plane assuming an unperturbed longitudinal motion.}}
\label{fig:asimm2}
\end{figure}

The Talbot-Lau interferometer, sketched and described in Fig.~\ref{fig:asimm2} can operate on incoherent uncollimated beams. This result is known as the Lau effect \cite{Jahns1979263,lau} and originates from the matching of the periodicity of the coherence function generated by the first grating, acting as a pure intensity mask, with the period  of the second grating, $d_2$ \cite{gori}. For a general discussion of the coherence properties of particle beams and the coherence requirements of different interferometers see \cite{cronin_coherence}. For our current purposes, it is sufficient to recall that the intensity pattern produced by a fully incoherent beam can be modeled by integrating the intensity distribution of point sources placed on the plane of the first grating \cite{concepts, amattint,patorski}. Schematically, the intensity pattern measured at the detection plane is given by
\begin{equation}
I(x|\lambda) = \int I_{\rm{Point}}(x|x_0,\lambda) |\mathcal{T}_1(x_0)|^2 dx_0
\label{eqn:pointsource}
\end{equation}
where $I_{\rm{Point}}(x|x_0,\lambda)$ is the intensity pattern produced by a monochromatic point source of wavelength $\lambda=\lambda(v)$ illuminating the second grating from the point $x_0$. The function $\mathcal{T}_1(x)$ is the transmission function of the first grating \cite{goodman}. In the case of non-monochromatic beams (as considered in section \ref{sec:Monte_Carlo}), the intensity pattern $I(x)$ is found by further integrating $I(x,\lambda)$ weighted by the probability distribution $p_{\lambda}(\lambda)$, or equivalently the speed distribution $P(v)$:
\begin{equation}
I(x)_{\rm{NM}} = \int I_{\rm{Point}} \left( x|x_0, \lambda  \right) |\mathcal{T}_1(x_0)|^2  P(v) dx_0 dv
\label{eqn:intensity_NM_gen}
\end{equation} 

The Fresnel integral formalism cannot easily take into account the presence of an external force acting on the interfering particles. We will thus employ an equivalent description of the Talbot-Lau interferometer based on the Wigner function \cite{CaseWigner, LeeWigner}. This approach allows to incorporate a constant acceleration $a$ in a straightforward manner \cite{arndtonatomint, CaseWigner}, and has been used to obtain the intensity pattern for symmetric configurations ($d_1 = d_2 $ and $\eta=1$) in the presence of a transverse acceleration \cite{arndtonatomint}. The same theoretical framework has also been applied to the asymmetric setups of our interest ($d_1 \neq d_2 $ and $\eta \neq 1$), but the external force was neglected \cite{beyondeikonal}. We now present a general calculation that takes both effects into account.

Following \cite{arndtonatomint} and  \cite{beyondeikonal}, we introduce the Wigner function phase-space representation of the quantum state of the particle within the interferometer, given its density operator $\rho$:
\begin{equation}
W(x,p) = \frac{1}{2 \pi \hbar} \int \e^{i p s/\hbar} \bra{ x - s/2}  \rho  \ket{ x + s/2} ds .
\label{eqn:wigner}
\end{equation}
We recall that the relevant degrees of freedom are the \emph{transverse} center-of-mass position and momentum ($x$ and $p$ respectively), whereas the longitudinal motion (along the $y$-axis in Fig.~\ref{fig:asimm2}) is assumed to be essentially classical. Specifically, uniform motion satisfying $t=y/v$ where $v$ is the longitudinal speed of the particle. This relation links time to the longitudinal space evolution of the interference pattern.
\corr{Wigner representation turns out to be very useful when the Hamiltonian is at most quadratic in the position and momentum operators, since, in this case, its time evolution has a simple analytical expression \cite{CaseWigner,LeeWigner}}. For example for the case of a linear potential of the form $V(x) = -x \, m a$, that is one resulting in a constant acceleration $a$ along the $x$-direction, the Wigner function evolved at time $t$ given the initial state $W_0(x,p)$ at $t=0$, reads \cite{arndtonatomint,beyondeikonal}:
\begin{equation}
W_t(x,p) = W_0 \left( x - \frac{pt}{m} + a\frac{t^2}{2}, p-mat \right).
\label{eqn:freeevol}
\end{equation}
The intensity distribution at the time $t$, namely,  $I_t(x)$, is then given by the marginal distribution
$$
I_t(x) = \int  W_t(x,p) dp.
$$
To describe a Talbot-Lau interferometer, the transformation induced on the Wigner function by the interaction with the gratings is needed. We assume that the action of the latter on the incoming single-particle state \corr{$\rho=\ket{\psi}\bra{\psi}$}, represented by the wavefunction $\psi(x) = \braket{x | \psi}$, can be modelled by a transmission function $\mathcal{T}(x)$. The wavefunction after the grating is then $\psi'(x) = \mathcal{T}(x) \psi(x)$, and the corresponding transformation on the Wigner function reads:
\begin{align}
\widetilde{W}(x,p) &= \frac{1}{2 \pi \hbar} \int ds \, \e^{i p s / \hbar} \, \mathcal{T}(x-s/2)\, \mathcal{T}^*(x+s/2) \nonumber \\ 
&\hspace{2.5cm}\times \braket{ x - s/2 | \psi} \braket{\psi  | x + s/2}, \\
&= \int dx_0 dp_0 K(x,x_0; p,p_0) W(x_0,p_0),
\label{eqn:gratingevol}
\end{align}
where $K(x,x_0; p,p_0)$ is defined as \cite{beyondeikonal}:
\begin{align}
K(x,x_0; p,p_0) &= \frac{\delta(x-x_0)}{2 \pi \hbar} \nonumber\\
&\hspace{-0.5cm} \times \int ds	\, \e^{i (p-p_0 ) s / \hbar } \,\mathcal{T}(x-s/2) \mathcal{T}^*(x+s/2)  .
\label{eqn:gratingpropagator}
\end{align}

\corr{Assuming a complete incoherence of the incoming particle beam, namely, $\Delta p \gg \hbar / d_1$ \cite{beyondeikonal}, where $\Delta p$ is the variance of the transverse momentum distribution, the corresponding Wigner function after the grating $G_1$, defined by its transmission function $\mathcal{T}_1(x)$, reads \cite{beyondeikonal}:}
\begin{equation}
\widetilde{W}_0(x,p) = \frac{1}{\mathcal{N} p_y} |\mathcal{T}_1(x)|^2 \mathcal{P} \left( \frac{p}{p_y} \right),
\label{eqn:initialwigner}
\end{equation}
where we denoted with $p_y = m v$ the longitudinal momentum, $\mathcal{P}(p/p_y)$ is the transverse momentum distribution, and $\mathcal{N}$ is a suitable normalization factor.

The initial state \eqref{eqn:initialwigner}, first undergoes free evolution for a time $T_1 = L/v$, governed by equation \eqref{eqn:freeevol}. The grating transformation \eqref{eqn:gratingevol} is then applied with the transmission function $\mathcal{T}_2$ of $G_2$, followed by free evolution for a time $T_2 = \eta T_1$ to obtain the Wigner function at the detection plane: $W_{T_1 + T_2}(x,p) \equiv W_2(x,p)$ (see Appendix \ref{app:B} for the explicit calculation). 

\corr{Upon defining the Fourier expansions of the two functions of $G_1$ and $G_2$, namely:
\begin{equation}
|\mathcal{T}_1(x)|^2 = \sum_{l=- \infty}^{\infty} A_l\, \e^{i 2 \pi l x / d_1}
\label{eqn:transmission1}
\end{equation}
and
\begin{equation}
\mathcal{T}_2(x) = \sum_{n=- \infty}^{\infty} b^{(2)}_n\, \e^{i 2 \pi n x / d_2},
\label{eqn:transmission2}
\end{equation}
the intensity distribution $I(x) \equiv  \int W_2(x,p) dp$ reads:
\begin{equation}
I(x) =\frac{1}{\mathcal{N}} \sum_{l=-\infty} ^{\infty} A^*_l B_{l \cdot q} \left( \alpha_l \right)
\exp \left\lbrace \frac{2 i \pi l}{\eta d_1} \left[ x - \Delta x \right] \right\rbrace.
\label{eqn:finalresult}
\end{equation}
where %$B_k(\alpha)$ (with $k=l \cdot q$) are the so called \emph{Talbot coefficients} \cite{arndtonatomint} defined as: 
\begin{equation}
B_{l \cdot q} (\alpha_l) = \sum_{n= -\infty} ^{\infty} b^{(2)}_n (b^{(2)} _{n-l \cdot q})^* \, \e^{i \pi \alpha_l (l \cdot q - 2n)}.
\label{eqn:talbotcoeff}
\end{equation}
are the so-called \emph{Talbot coefficients} \cite{arndtonatomint} with:
\begin{equation}
\alpha_l=  \frac{L}{L_T} \frac{d_2}{d_1} l.
\label{eqn:alpha}
\end{equation}
which contains the usual definition of the Talbot length $L_T = d_2^2 / \lambda$ \cite{patorski, clauser}.
Notice that the parameter $q$ must be an integer number, and reads:
\begin{equation}
q=\frac{d_2}{d_1} \frac{(1+ \eta)}{ \eta},
\label{eqn:integer}
\end{equation}
}
Finally, it is apparent that the effect of a nonzero acceleration $a$ is to rigidly displace the fringe pattern by the following quantity:
\begin{equation}
\Delta x = a \frac{T_1^2}{2} \eta(\eta+1),
\label{eqn:deltax}
\end{equation}
which is proportional to $a T_1^2$ as anticipated. We fully discuss the inertial displacement in section \ref{sec:inertial_sensitivity}.

Being based on the transmission function formalism, this model is very general and can be applied to a wide range of particles and diffraction gratings at $G_2$, pure intensity masks as well as phase gratings that alter the phase of the incoming wavefunction. Furthermore, this treatment can also account for a broad range of particle-gratings interactions; examples include the van der Waals atom-surface interaction \cite{amattint, vdw,savas2} or electrostatic forces for charged particles \cite{amattint, savas}.  

Sufficiently weak interactions in particular result in a reduced \emph{effective} slit width \cite{amattint, vdw}, that has also been observed experimentally \cite{arndt}. For stronger interactions, a more general approach beyond the Eikonal approximation \cite{beyondeikonal} can still make use of this formalism. The properties of the general equation \eqref{eqn:finalresult} are now discussed in detail for the cases of our interest.

\subsection{Features of the interference pattern and resonance conditions }

First of all we note that equation \eqref{eqn:finalresult}, describing the statistical interference fringe pattern in a Talbot-Lau interferometer, is a Fourier series expansion with a magnified period $d_3 \equiv \eta d_1$. The dependence on the length $L$ only enters through the dimensionless ratio $L/L_T$. The factor $\eta$ can also be less than unity, however we are particularly interested in the case $\eta >1$, therefore, from now on we will refer to $\eta$ as the \emph{magnification factor}. As we mentioned, the properties of the gratings are encoded in the coefficients $B_{l \cdot q}(\alpha_l)$ and $A_l$. For the sake of clarity, we now specialise our analysis to gratings described by the following (single period) transmission function: 
$$
\mathcal{T}(x|w,z,f,d)=
\left\{
\begin{array}{ll} w & \mbox{if} \quad x \in [0,fd]\\
z & \mbox{if} \quad x \in [fd,d],
\end{array}
\right.$$
where $w,z \in \mathbb{C}$ are two complex numbers, $d$ is the grating period, and $f$ the \emph{open fraction} of the grating. This form is particularly convenient since, upon writing the Fourier expansion $\mathcal{T}(x,|w,z,f,d) = \sum_n b_n(w,z,f) \exp \left\lbrace i 2 \pi x/d \right\rbrace$, we have the following analytical expression for the Fourier coefficients:
\begin{equation}
b_n(w,z,f) = f\sinc(\pi n f) \left(z \,\e^{- i \pi n f} + w \,\e^{ i \pi n f}\right).
\label{eqn:fouriercoeff}
\end{equation}
Partial transparency of the grating substrate together with a possible (constant) phase added could be accounted for by a suitable choice of $w$ and $z$. 
However, in the rest of this paper we set $w=1$ and $z=0$, to describe material gratings realised as open slits in a substrate \cite{reticoli, arndt}. The open fraction then corresponds to the ratio between the slit width and the grating period. Furthermore, in the following we drop the explicit dependence of the Fourier coefficients \eqref{eqn:fouriercoeff} on the parameters $(w,z,f)$ as the two gratings $G_1$ and $G_2$ are assumed to have the same open fraction and transmission properties. 

\begin{figure*}[t!]
%\begin{minipage}[t]{0.48\linewidth}
\includegraphics[width=.7\columnwidth]{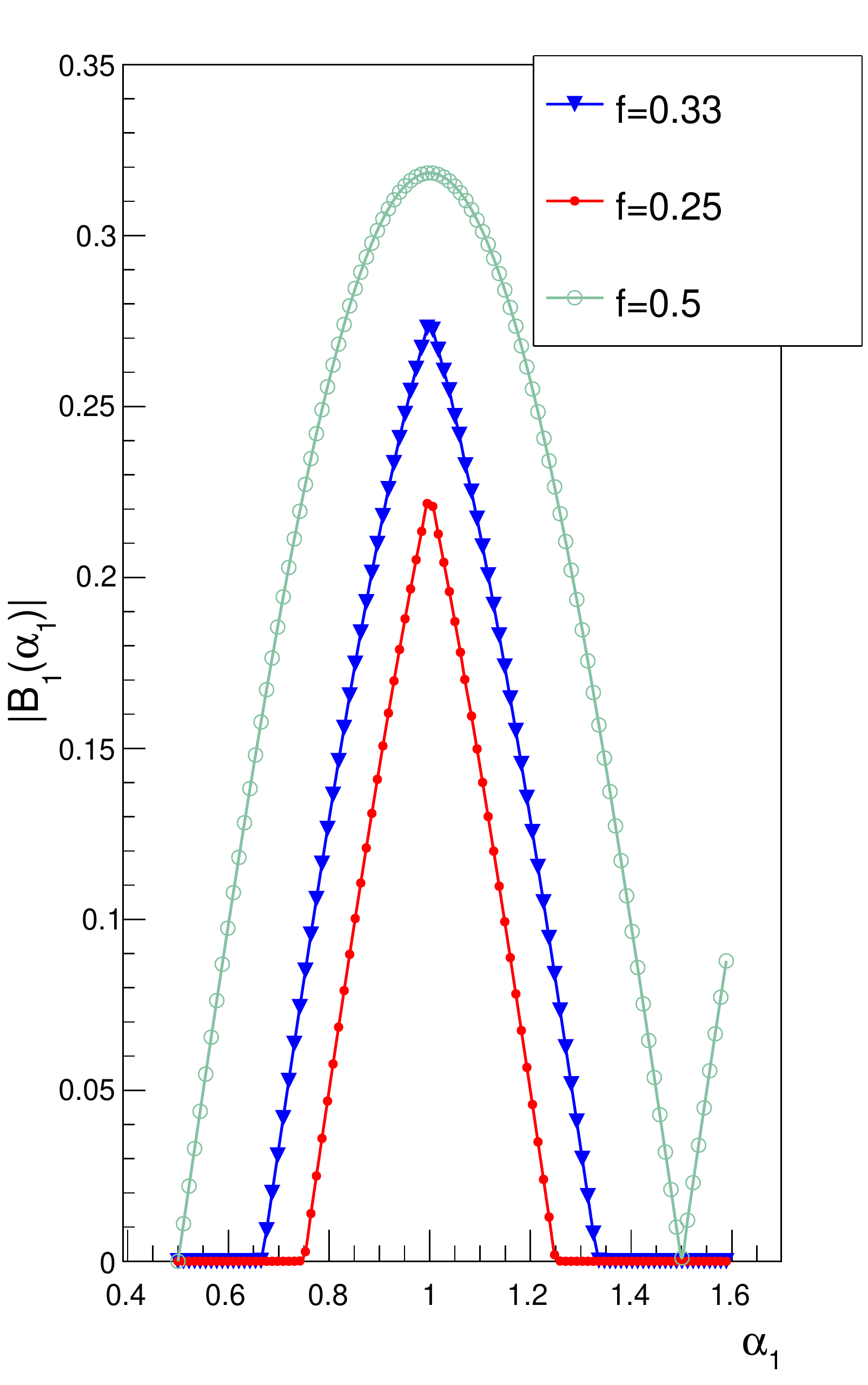}
\hspace{2.5cm}
%\end{minipage}\hfill%
%\begin{minipage}[t]{0.48\linewidth}
\includegraphics[width=.7\columnwidth]{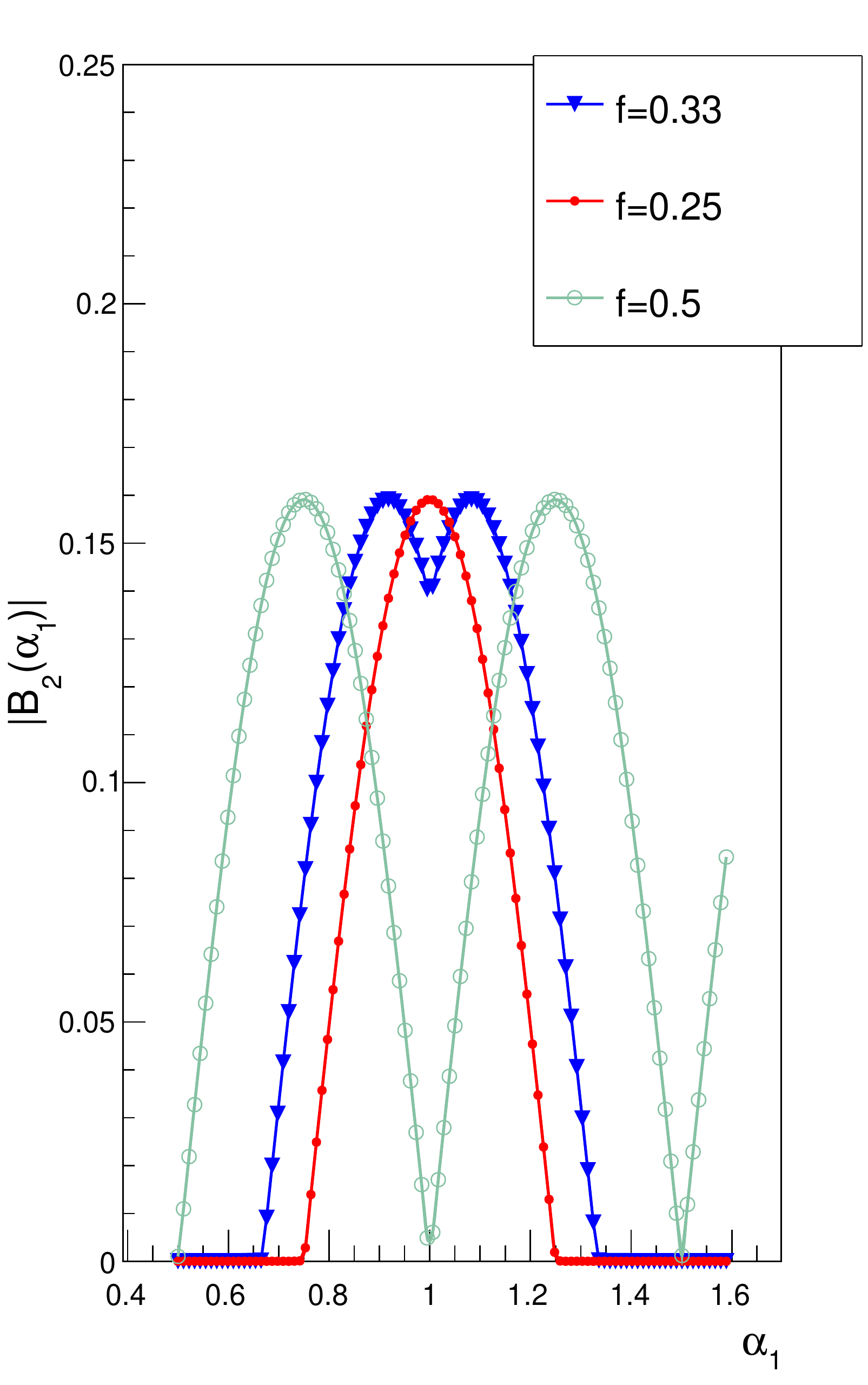}
%\end{minipage}%
 \vspace{-0.3cm}
\caption{Plots of the functions $|B_1(\alpha_1)|$, on the left, and $|B_2(\alpha_1)|$, on the right, for three values of the open fraction $f=0.33,0.25,0.5$. This results have been calculated by truncating the summation of equation \eqref{eqn:talbotcoeff} to $|n|<20$. It is apparent that the shape of the functions $|B_q(\alpha_1)|$ strongly depends on the open fraction, and is always symmetrical with respect to $\alpha_1=1$. The functions are also periodic, reflecting the properties of the underlying Talbot effect, however, only one period is shown for the sake of clarity. }\label{fig:talbotcoeffplots}
\end{figure*}
Now we look for the resonance conditions of equation \eqref{eqn:finalresult}, i.e., the set of parameters $\eta, d_1, d_2, L/L_T$ that maximises the visibility of the pattern. We recall that the visibility or contrast $C$ of the fringe pattern $I(x)$ is defined as 
\begin{equation}
C = \frac{I_{\rm max} - I_{\rm min} }{I_{\rm max} + I_{\rm min}}.
\label{eqn:visibilitydefinition}
\end{equation}
Since the function $I(x)$ is a Fourier series, one can truncate the summation to the lowest orders, and consider the visibility of the resulting sinusoidal function as a good approximation of the actual visibility. This parameter is called the \emph{sinusoidal visibility} \cite{arndt,arndt2}, and for equation \eqref{eqn:finalresult} it reads:
\begin{equation}
C_{\rm sin}(\alpha_1,q) = 2 \frac{|A_0 B_q(\alpha_1)|}{|A_0|^2}  =  2 \frac{| B_q(\alpha_1)|}{|A_0|}.
\label{eqn:sinusoidal}
\end{equation}
The constant coefficients $A_0$ are the zeroth-order Fourier coefficients of $|\mathcal{T}_1(x)|^2$, the intensity transmission function of the first grating (for the case $z=1$ and $w=0$ it coincides with the transmission function itself). Equation \eqref{eqn:sinusoidal} suggests the modulus of the $q$-th Talbot coefficient as a good estimator of the pattern visibility. The requirement that $q$, defined in \eqref{eqn:integer}, is an integer allows to enumerate different families of \emph{resonance conditions} as a function of the physical parameters, in particular we focus on the following  choices:
\begin{subequations}
\label{eqn:resonance_2}
\begin{align}
q=2  \Rightarrow  &\frac{d_1}{d_2} =  \frac{(1 + \eta)}{2 \eta} \quad \mbox{(symmetric)}, \label{eqn:resonance_2_symm}\\
q=1  \Rightarrow  &\frac{d_1}{d_2} = \frac{(1 + \eta)}{\eta} \quad \mbox{(asymmetric)}. \label{eqn:resonance_2_asymm}
\end{align}
\end{subequations}
We can see that the magnification factor determines the ratio of the two grating periods. The most common standard Talbot-Lau setup (that we will refer to as the \emph{symmetric setup}) belongs to the case $q=2$, and has $\eta=1$, implying that $d_1=d_2$. In the following we will study the interesting properties of the \emph{asymmetric setup} with $q=1$ and $\eta>1$. 
\par
The value of $q$ determines the relevant Talbot coefficients influencing the visibility, respectively $B_1(\alpha_1)$ for the asymmetric case, and $B_2(\alpha_1)$ for the standard symmetric setup. We now turn our attention to the $\alpha$ dependence of $|B_1(\alpha_1)|$ and $|B_2(\alpha_1)|$. The two functions are plotted in Fig.~\ref{fig:talbotcoeffplots} for different values of the open fraction $f$.
The position of the relative maximum of the relevant Talbot coefficient sets the \emph{resonance} condition on the length. For instance we see that for the $B_1(\alpha_1)$, this always occurs for $\alpha_1=1$, whereas the behaviour of $B_2(\alpha_1)$ is more irregular and depends on the open fraction. Assuming for definiteness that the maximum occurs for $\alpha_1 =1$, and using the definition \eqref{eqn:alpha}, we obtain:
\begin{equation}
L= \frac{d_1}{d_2} L_T = \frac{d_1 d_2}{\lambda},
\label{eqn:resonance_1}
\end{equation}
where the periods of the gratings and magnification factor $\eta$ have to satisfy either of the conditions \eqref{eqn:resonance_2} (or any other combination corresponding to an integer value of $q$, defined by \eqref{eqn:integer}).  
Quantum diffraction takes place at the second grating, so in this general configuration with $d_1 \neq d_2$, it is $d_2$ that sets the relevant length scale trough the Talbot length $L_T = d_2^2 / \lambda $. Furthermore, as a manifestation of the underlying Talbot effect, resonance is possible also at higher integer multiples of $L_T$. This is reflected in the periodicity of the Talbot coefficients in their argument $\alpha_1$.

The case of a symmetric setup ($q=2, \eta=1, d_1=d_2$) with $f=0.5$ is peculiar, and does not satisfy the same resonance conditions, since is is evident from Fig.~\ref{fig:talbotcoeffplots} that it achieves a maximum visibility for $\alpha \neq 1$. The case $f=0.5$ is also critical in the classical case: the visibility of a classical moiré deflectometer with $f=0.5$ is exactly zero \cite{moiree}. It is interesting to see that if an asymmetric setup is employed, all the chosen values of open fraction, including $f=0.5$,  behave similarly. We will analyse the consequences of this property in section \ref{sec:Monte_Carlo}.

\subsection{Asymmetric setups and period magnification} \label{sec:asymmetric_setups}

\begin{figure}[htbp]
\centering
 \includegraphics[width=0.35\textwidth]{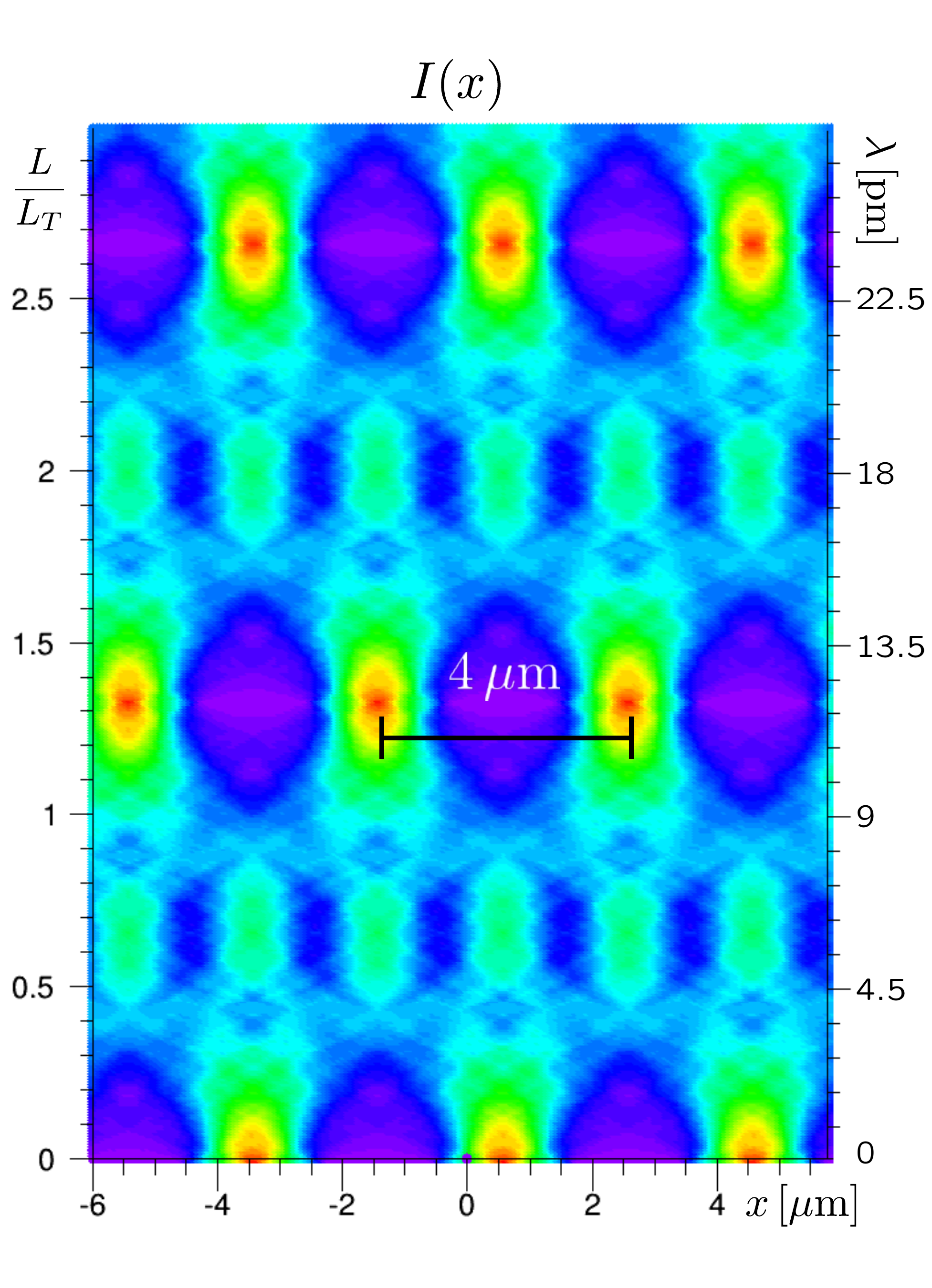}
 \vspace{-0.5cm}
\caption{Plot of the intensity $I(x)$, from equation \eqref{eqn:finalresult}, also as a function of the de Broglie wavelength which varies along the $y$-axis. For definiteness we set realistic parameters for low energy electrons: $d_2 = 1 \unit{\mu m}$, $d_1 = 4/3\, d_2$, $\eta=3$,  $L=0.11 \unit{m}$ and $f=0.3$. This choice satisfies the asymmetric resonance conditions for $10 \unit{keV}$ electrons. As predicted we see the main interference fringes appear at $L/L_T = d_1 / d_2$ and have a magnified period $d_3 = \eta d_1$. We inserted real physical scales for clarity, but we note that only the adimensional ratios $L/L_T$ and $d_1 / d_2$ appear in \eqref{eqn:finalresult}, so our plot shows the general form of the \emph{Talbot carpet} for the $\eta=3$ configuration with $f=0.3$ material gratings.}\label{fig:carpet_a}
\end{figure}

\begin{figure}[htbp]
\centering
 \includegraphics[width=0.35\textwidth]{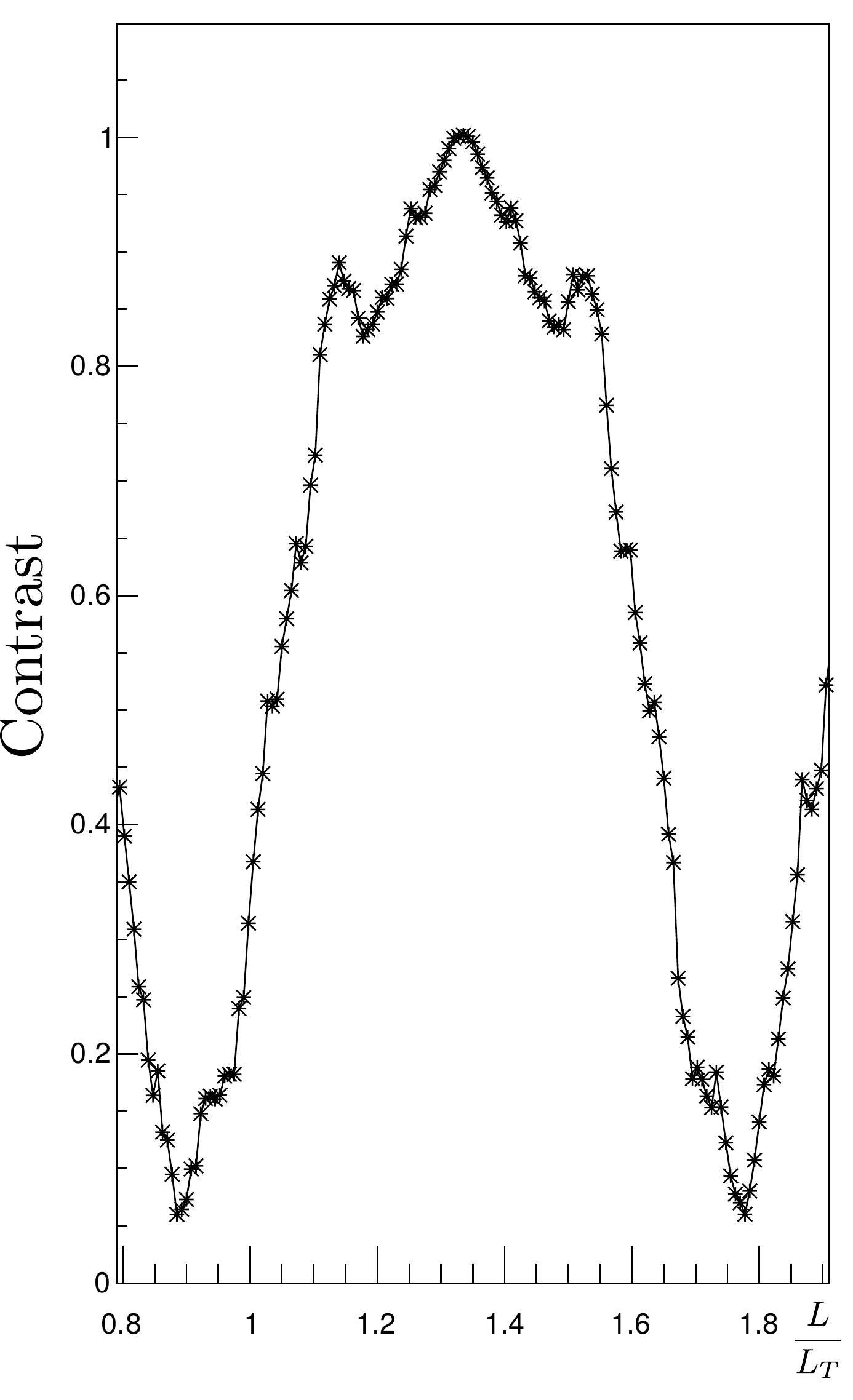}
  \vspace{-0.3cm}
\caption{Visibility of the interference pattern of Fig.~\ref{fig:carpet_a}, in the neighborhood of the main interference peak, calculated with equation \eqref{eqn:visibilitydefinition}. The maximum occurs at $L/L_T = d_1/d_2$, as predicted in section \ref{sec:general_description}. Note that the shape of the peak differs from the plots of Fig.~\ref{fig:talbotcoeffplots}. Those curves are proportional to the sinusoidal visibility, which is only an approximation, while here there are contributions from higher orders of the Fourier series \eqref{eqn:finalresult}. }\label{fig:carpet_b}
\end{figure}

In order to study the effect of the asymmetric configuration on the interference pattern, it is useful to start from a specific example. Choosing the resonance condition \eqref{eqn:resonance_2_asymm}, interference pattern is given by the general equation \eqref{eqn:finalresult}. The relevant properties of the interference patterns can be summarised in a \emph{carpet} as shown in Fig.~\ref{fig:carpet_a}. \corr{This is a two dimensional density plot where each section is the intensity distribution $I(x)$ for a given value of $L/L_T$: the carpet can be scanned by tuning the particle energy (or the de Broglie wavelength) to adjust $L_T$. The behaviour of the visibility is shown in Fig.~\ref{fig:carpet_b}.}

In general, the features of the asymmetric Talbot-Lau setup can be described as follows:
\begin{itemize}
\item The maximum fringe period is magnified and given by $d_3 = \eta d_1$. Fractional revivals are also present and are peculiar of the Talbot effect (see Fig.~\ref{fig:carpet_a}).
\item The total length is given by $L^{(\mathrm{TOT})} = L(1+ \eta)$. Imposing the appropriate resonance conditions on the grating periods \eqref{eqn:resonance_2_asymm} yields:
\end{itemize}
\begin{equation}
L^{(\mathrm{TOT})} = (1+ \eta) \frac{d_1}{d_2} L_T = (1 + \eta) \frac{d_1 d_2}{\lambda} 
\label{eqn:ltot}
\end{equation}
The properties of the two configurations relevant for the calculations to follow are summarised in Table \ref{tab:table1}.

\renewcommand{\arraystretch}{2.5}
\newcolumntype{C}[1]{>{\centering\arraybackslash}p{#1}}
\begin{table}[h!]
\centering
\begin{tabular}{c | C{2cm} | C{2cm}}
 & \textbf{Symmetric} & \textbf{Asymmetric} \\
\hline \hline 
 $ \displaystyle L^{(\rm{TOT})}  $ & $\displaystyle 2\frac{d_{2,\rm{s}}^2}{\lambda} $ & $\displaystyle \frac{(\eta + 1)^2}{\eta} \frac{d_{2,\rm{a}}^2}{\lambda} $ \\
\hline 
$\displaystyle d_3  $ & $ \displaystyle d_{2,\rm{s}}$ & $ \displaystyle (\eta+1) d_{2,\rm{a}} $ \\
\hline \hline
\end{tabular} 
\caption{\label{tab:table1}  Summary of the relevant properties $L^{(\rm{TOT})}$ and  $d_3 = \eta d_1$ for the symmetric setup \eqref{eqn:resonance_2_symm} with $d_1=d_2=d_{2,\rm{s}}$, and the asymmetric setup \eqref{eqn:resonance_2_asymm} with $d_2=d_{2,\rm{a}}$ and $d_1=(\eta+1) d_{2,\rm{a}}/\eta$. }
\end{table}
It is possible to prove that for a given energy (wavelength) and at a fixed total interferometer length, asymmetric configurations allow to maximize the period of the interference fringes with respect to the symmetric setup. If the ratio $r=d_{3,\rm{a}} / d_{3,\rm{s}}$ is evaluated under the constraint that the two interferometers are of the same total length, namely, 
\begin{equation}
\frac{d_{2,\rm{a}}^2}{d_{2,\rm{s}}^2} = \frac{2 \eta}{(\eta+1)^2},
\label{eqn:ratio}
\end{equation} 
the following result is obtained:
\begin{equation}
\left. r \right|_{\mbox{\tiny{Equal length}}}= \frac{d_{3,\rm{a}}}{d_{3,\rm{s}}} =  \eta \cdot \frac{\eta+1}{\eta} \frac{d_{2,\rm{a}}}{d_{2,\rm{s}}} = \sqrt{2 \eta}
\end{equation}
So we see that asymmetric configurations provide a systematic improvement of the ratio $d_3/L^{(\rm{TOT})}$ that scales well with the magnification factor. This can be of interest experimentally for a variety of cases \cite{amattint}. Magnifying configurations have been actually realized for low energy electrons \cite{lauint}, using however different resonance conditions and an extreme ($\eta=100$) magnification factor, so that the observation plane was effectively in the far field of the second gratings. As a matter of fact that configuration requires different coherence conditions than the Talbot-Lau interferometer and is referred to as a \emph{Lau interferometer} \cite{cronin_coherence}.

\section{Inertial sensitivity and applications} \label{sec:inertial_sensitivity}

Now we turn our attention on the inertial sensitivity of Talbot-Lau interferometers. In section \ref{sec:general_description}, we determined that the displacement of the pattern induced by an external acceleration $a$ is given by Eq. \eqref{eqn:deltax}. 

This is a generalisation of the result from \cite{arndtonatomint} that allows for gratings of different periods and a magnification factor $\eta$. If we set $\eta=1$  we obtain the  $\left. \Delta x \right|_{\eta=1} = a T_1^2$, which is the well known displacement law for the the geometrical shadow pattern in a moiré deflectometer due to the same effect \cite{moiree}. This correspondence will be further discussed, also in the asymmetric configuration, in section \ref{sec:moire}.

The displacement \eqref{eqn:deltax} is quadratic in the magnification factor $\eta$. This is an interesting property that might be of great help in those experimental situations where the total length of the setup is limited by the properties of the interfering particles. For example if they have a finite lifetime \cite{amattint}, or need to propagate in vacuum, under shielding from stray fields or in a cryogenic environment \cite{amattint, clauserli}. An interesting potential application for a Talbot-Lau inertial sensor is the measurement of the gravitational acceleration $g$ of the positronium atom. In this situation all the experimental complications we mentioned are in effect.

It is apparent that at a fixed total length $L^{(\mathrm{TOT})} = L(1+ \eta)$, increasing the asymmetry factor $\eta$ also reduces $T_1$, and the dependence on both parameters of $\Delta x$ is quadratic. For this reason we apply the same reasoning of section \ref{sec:asymmetric_setups} to find if there is a systematic gain in the inertial displacement from symmetric to asymmetric setups of the same length.

Now we evaluate the displacement per unit interferometer length, namely:
$$
r_{\Delta x} = \Delta x / L^{(\rm{TOT})}
$$ 
in the symmetric and asymmetric case under the constraint \eqref{eqn:ratio}. Using the same notation of section \ref{sec:asymmetric_setups}, one can prove that (assuming $\eta > 1$):
\begin{equation}
\frac{r_{\Delta x, \rm{a}}}{r_{\Delta x,\rm{s}}} = \left. \frac{\Delta x_{\rm{a}}}{\Delta x_{\rm{s}}} \right|_{\mbox{\tiny{Equal length}}} = \frac{2 \eta}{(\eta +1)} > 1
\label{eqn:ratio2}
\end{equation}
This factor is greater than unity, but it is limited to a maximum value of  $\Delta x_{\rm{a}} / \Delta x_{\rm{s}} =2$ for $\eta \gg 1$. However, already for $\eta=3$ one can magnify the \emph{fall} of the beam by $50 \%$ with respect to a symmetric configuration of the same length. This can already be a sizeable gain for some specific applications.

From a practical point of view, it can be proven \cite{Batelaan199785}, that the relative uncertainty $\sigma_a /a$ with which the acceleration $a$ can be measured by detecting a shift $\Delta x$ in a fringe pattern with period $d_3$ and contrast $C$ reads:
\begin{equation}
\frac{\sigma_a}{a} = \frac{1}{\sqrt{N}} \frac{1}{2 \pi C \, \Delta x / d_3 } ,
\label{eqn:sensitivity}
\end{equation}
where $N$ is the number of data points forming the pattern, which depends on the beam intensity and the efficiency of the detector. Furthermore, the contrast is mainly influenced by the longitudinal velocity spread of the incoming particles. For a non-monochromatic beam the intensity pattern is easily recovered by integrating over the speed distribution (see \cite{amattint} and references therein). The result is in general a loss in visibility that depends on the width of the velocity distribution (see section \ref{sec:Monte_Carlo}). The chosen Talbot-Lau configuration directly influences the inertial sensitivity via the \emph{relative displacement} $\Delta x / d_3$, where we recall that $d_3$ is the period of the interference fringes. It is thus useful to derive an expression for the ratio in the two cases of our interest. Starting from the asymmetric setup, defined by the resonance conditions \eqref{eqn:resonance_2_asymm} and  \eqref{eqn:resonance_1} we have
\begin{equation}
 \frac{\Delta x_{\rm{a}}}{d_{3,{\rm{a}}}} = \frac{a T_1^2 \eta (\eta +1)}{2 \eta d_{1,{\rm{a}}}} = \frac{a}{2} \frac{\sqrt{\eta}}{(\eta+1)} \sqrt{\frac{m}{h}} \left[  T_{(\rm{TOT})} \right]^{3/2}.
\label{eqn:relative_asymm}
\end{equation}
The last equality follows from simple substitutions and algebraic manipulations using equation \eqref{eqn:ltot}, the resonance conditions and the definitions of the Talbot length $L_T = d_2^2 / \lambda $ and of the de Broglie wavelength. We introduced $T_{(\rm{TOT})}$ for the total flight time from the first grating to the detection plane, namely $L^{(\rm{TOT})} /v$.

On the one hand, equation \eqref{eqn:sensitivity} tells that to improve the sensitivity, the relative displacement $\Delta x / d_3$ should be maximised. On the other hand, equation \eqref{eqn:relative_asymm} shows that this quantity increases monotonically with the total flight time, as expected, but also that it tends to zero as $\eta \gg 1$. 
\par
We can physically motivate the dependence of Eqs.~\eqref{eqn:relative_asymm} and \eqref{eqn:relative_symm} from the particle mass: for a fixed total interferometer length and longitudinal speed $v$, any particle subjected to the same acceleration $a$, will undergo the same transverse displacement, according to equation \eqref{eqn:deltax}. However, the heavier the particle, the smaller its de Broglie wavelength would be, thus leading to shorter periods for the two gratings and for the resulting fringe pattern. Following this line of reasoning, the ratio $\Delta x / d_3$ is expected to increase with the particle mass.
\par
The relative displacement for the symmetric setup instead reads:
\begin{equation}
\frac{\Delta x_{\rm{s}}}{d_{3,{\rm{s}}}} = a \frac{1}{2 \sqrt{2}} \sqrt{\frac{m}{h}} \left[  T_{(\rm{TOT})} \right]^{3/2}.
\label{eqn:relative_symm}
\end{equation}
First we remark that it does not coincide with the result of equation \eqref{eqn:relative_asymm} for $\eta=1$. This is a consequence of the fact that the two configurations belong to two different sets of resonance conditions with different relevant Talbot coefficients, as discussed in section \ref{sec:general_description}. As a matter of fact the ``asymmetric configuration with $\eta=1$" differs from the standard symmetric setup because $d_1=2 d_2$ in the former, whereas $d_1=d_2$ in the latter; as required by \eqref{eqn:resonance_2_asymm}. Even in this case we have $ \Delta x_{\rm{a}} / d_{3,{\rm{a}}} < \Delta x_{\rm{s}} / d_{3,{\rm{s}}} $ at the same total length, so we see that the asymmetric setups we studied always provide a smaller relative displacement in the presence of a constant acceleration. This property, according to equation \eqref{eqn:sensitivity} can be a disadvantage if the aim is to measure the acceleration $a$ with great accuracy. However, in some realistic experimental situations it may be preferable to have a larger absolute displacement at the expense of the relative shift (for instance, due to the finite detector resolution).
\par
\begin{figure}[htb]
 \includegraphics[width=0.45\textwidth]{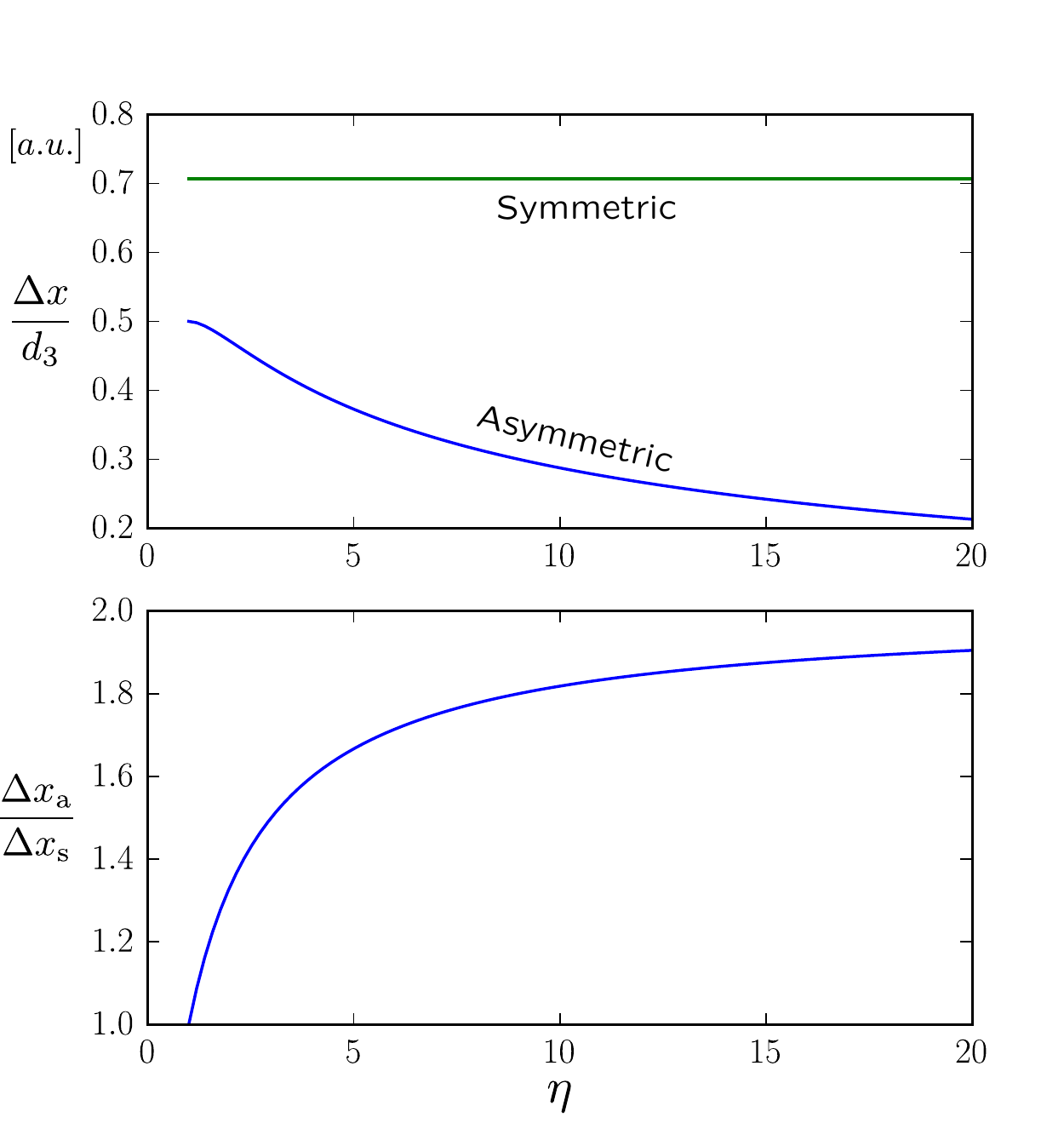}
 \vspace{-0.3cm}
\caption{The top panel shows a comparison of Eqs.~\eqref{eqn:relative_asymm} and \eqref{eqn:relative_symm} (assuming arbitrary units in which $a=h=m=T^{(\rm{TOT})} =1$). It is apparent that the asymmetric setup always provides a smaller relative displacement under the acceleration $a$. The absolute displacement is, however, always larger, as shown in the bottom panel. A smaller relative displacement is not desirable for inertial measurements, but also implies a reduced sensitivity to external disturbances. Generally the best trade off has to be found depending on the specific experimental application.}\label{fig:plots_scaling}
\end{figure}
A comparison of Eqs.~\eqref{eqn:relative_asymm} and \eqref{eqn:relative_symm} suggests that the impact of random external perturbations on the pattern visibility is effectively reduced by using a magnifying setup of the same length.
An example is the Lorentz force acing on charged particle due to stray electromagnetic fields \cite{amattint}. To summarise the results of this section, Figure \ref{fig:plots_scaling} displays the scaling with $\eta$ of the parameters we studied.  

\subsection{Comparison with moiré deflectometers} \label{sec:moire}
A moiré deflectometer, as described in \cite{moiree}, is a two-grating setup completely analogous to the one shown in Fig.~\ref{fig:asimm2}. The crucial difference is that the grating periods $d^{(\rm{m})}$ and the length $L^{(\rm{m})}$ are chosen to satisfy the constraint \cite{amattint}:
\begin{equation}
L^{(\rm{m})} \frac{\lambda}{d^{(\rm{m})}} \ll d^{(\rm{m})} \rightarrow L^{(\rm{m})} \ll L_T.
\label{eqn:classicality}
\end{equation}
Therefore, quantum diffraction is negligible. We introduced the superscript $(\rm{m})$ to denote the grating periods and length of the classical configuration. 
 Equation \eqref{eqn:classicality} implies that a given resonant Talbot-Lau setup with parameters $\eta, L, d_1, d_2$ at fixed de Broglie wavelength $\lambda$, can be made into a classical device by changing the grating periods to larger values: $d_i^{(\rm{m})} \gg d_i$ for $i=1,2$. On the other hand one could decrease the length and keep the same gratings, so that $L^{(\rm{m})} \ll L$. However, since we are interested in inertial sensing application and the fringe displacement strongly depends on the length, in the following we always assume that the first route is taken when comparing the two devices. 

For this reason, a moiré deflectometer will always produce a fringe system with a larger period than the Talbot-Lau setup of the same length tuned for the same particle beam. 

We now derive in very simple terms the main features of the classical fringe pattern in the presence of an external acceleration $a$. Let us suppose that the incoming particle with speed $v$ starts with a transverse position and speed $(x_0, v_0)$, on a plane located at a distance $L_s = v T_s $ before the first grating. From the laws of uniformly accelerated motion it is then straightforward to write the following system of equations: 
\renewcommand{\arraystretch}{1.5}
\begin{equation}
\left\{\begin{array}{ll}
             x_1 &= x_0 + v_0 T_s + \frac{1}{2} a T_s^2\\
             x_2 &= x_0 + v_0 (T_1+T_s) + \frac{1}{2} a (T_1 + T_s)^2\\
             x_3 &=  x_0 + v_0 (T_1+ T_2 + T_s) + \frac{1}{2} a (T_1 + T_2 + T_s)^2 .
\end{array}\right. 
            \label{eqn:lawsofmotion}
\end{equation} 
Where  $x_1$, $x_2$ , $x_3$ are the $x$-positions of the particle on the plane of $G_1$, $G_2$ and the detector respectively, $T_1$ and $T_2$ being the corresponding times of flight. After some algebraic manipulations we can eliminate the dependence on $x_0$ and $v_0$, solving for $x_3$ as a function of $x_2$ and $x_2$:
\begin{equation}
x_3 = x_2 \left( 1 + \frac{T_2}{T_1} \right) - x_1 \frac{T_2}{T_1} + \frac{1}{2} a \left( T_2^2 + T_1 T_2 \right).
\label{eqn:finposmoire}
\end{equation}
It is worth noting that equation \eqref{eqn:finposmoire} does not depend on the initial conditions $x_0$ and $v_0$: only the dynamics after the first grating are relevant. The same expression could have been obtained by assuming initial conditions on the plane of $x_1$. One also sees that the displacement due to $a$ is the sum of two contributions depending on both times of flight, as expected since the force acts in both regions. 

Equation \eqref{eqn:finposmoire} must be coupled with the requirement that the intermediate arrival positions onto the gratings are contained in the support of the gratings transmission function. To get an intuitive picture we implement this requirement by the simple replacements
\begin{equation}
x_1 = n d^{(\rm{m})}_1 \quad \mbox{and} \quad x_2 = m d^{(\rm{m})}_2,
\label{eqn:replacement_moire}
\end{equation}
that constrain the $x$-positions to be exact multiples $n$ and $m$, respectively, of the grating periods. This substitution, together with $T_2 = \eta T_1$ yields:
\begin{equation}
x_3 = m  d^{(\rm{m})}_2 ( 1 + \eta) - n d^{(\rm{m})}_1 \eta + \frac{a T_1^2}{2} \eta (\eta +1 ).
\label{eqn:quasi_finfinposmoire}
\end{equation}
A physically interesting periodic pattern arises if the grating periods and $\eta$ are chosen to cast equation \eqref{eqn:quasi_finfinposmoire} in the form: $x_3 = b d^{(\rm{m})}_3  +\frac12 a T_1^2 \eta (\eta +1 ) $, where $b$ is an integer number that depends on $m,n$ and $d^{(\rm{m})}_3$ is the period of the fringes, generally depending on $\eta$. For example, the standard moiré deflectometer, defined by $d^{(\rm{m})}_2 = d^{(\rm{m})}_1$ and $\eta =1$ is a suitable choice. However, we observe that also by using the asymmetric resonance conditions \eqref{eqn:resonance_2_asymm} we obtain the following expression
\begin{equation}
x_3 = \eta d^{(\rm{m})}_1 \left(m-n \right) + \frac{a T_1^2}{2} \eta (\eta +1 ),
\label{eqn:finfinposmoire}
\end{equation}
meaning that the final position is a multiple of $d^{(\rm{m})}_3 = \eta d^{(\rm{m})}_1 $. As we anticipated, the last line shows that period magnification and $a$-dependent fringe displacement have the exact features in the classical and quantum description of the setup of Fig.\ref{fig:asimm2}. Given this similarity, all the considerations made about the sensitivity (see equation \eqref{eqn:sensitivity}) remain valid. An important remark is that all other parameters being equal, the requirement $d_i^{(\rm{m})} \gg d_i$ causes the classical configuration to always produce a smaller relative displacement $\Delta x / d_3$, thus generally lowering the sensitivity \eqref{eqn:sensitivity}. 

However, the properties of the quantum and classical fringe patterns are markedly different. For example in the moiré deflectometer the visibility is independent of the particle energy \cite{amattint}, as also shown in section \ref{sec:Monte_Carlo}. This is why we carefully referred to the output of the moiré deflectometer as \emph{geometrical shadow} patterns, in contrast with the genuine quantum interference fringes of a Talbot-Lau interferometer.

\section{Numerical analysis} \label{sec:Monte_Carlo}

In the Talbot-Lau interferometer, the parameters $C$ and $\Delta x$ appearing in equation \eqref{eqn:sensitivity} for the inertial sensitivity, strongly depend on the longitudinal speed distribution $P(v)$ of the particle beam. This section is devoted to a numerical analysis of this dependence.

For definiteness, we will assume that the function $P(v)$ is a \corr{Gaussian with variance $\sigma_v^2$}. Hence we can write the general expression:
\begin{equation}
\left( \frac{\sigma_a}{a} \right)_{\rm{NM}} \left(\sigma_v \right) = \frac{1}{\sqrt{N}} \frac{1}{2 \pi \, C(\sigma_v)  \, \Delta x _{\rm{eff}} (\sigma_v) / d_3 } ,
\label{eqn:sensitivityNM}
\end{equation}
\corr{where $C(\sigma_v)$ is the contrast, $ \Delta x _{\rm{eff}} (\sigma_v)$ is an effective displacement and the subscript $\rm{NM}$
recalls that it applies to non-monochromatic beams.}
As Eqs.~\eqref{eqn:finalresult} and \eqref{eqn:intensity_NM_gen} suggest, the intensity for a non-monochromatic beam in the presence of an external force has the general structure:
\begin{subequations}
\begin{align}
I_{\rm{NM}}(x)  &= \int    I(x - \Delta x(v)|v)  P(v) dv  \label{eqn:intensity_nm_talbot} \\ 
            & \approx  \int    I(x - \Delta x _{\rm{eff}}|v)  P(v) dv \label{eqn:intensity_nm_fit}
\end{align}
\end{subequations}
where $\Delta x(v)$ and $I(x|v)$ are given by Eqs.~\eqref{eqn:deltax} and \eqref{eqn:finalresult} respectively, and we highlighted the parametric dependence on $v$ for clarity. Since there is a dependence on the integration variable both from the argument and in the functional form of $I(x)$, the second equality is in principle an approximation.

The effective displacement $\Delta x_{\rm{eff}}$ we just introduced, is what contributes to the sensitivity of the apparatus in Eq.~\eqref{eqn:sensitivityNM}, and depends on $\sigma_v$ (see Appendix \ref{app:C} for more details). The intensity factor $N$, can always be defined as $N={\cal M} T_{\rm{int}}$, where ${\cal M}$ is the beam intensity at the detector and a $T_{\rm{int}}$ is the integration time. As we will show, the visibility of a Talbot-Lau pattern is very sensitive to the $v$-distribution, so in many realistic particle beams, a velocity selection could be needed. In these cases the factor ${\cal M}$ (and in turn $N$) depends on $\sigma_v$ as well, and the best trade-off between visibility (decreasing with $\sigma_v$) and statistics (increasing with $\sigma_v$) has to be found. Since this study is specific to each experimental situation, in the following we only focus on the functions $C(\sigma_v)$ and $\Delta x_{\rm{eff}} (\sigma_v)$.
\subsection{Results}
Numerical integration of equation \eqref{eqn:intensity_nm_talbot} has been performed on a discrete set of points, with a standard normal speed distribution 
\begin{equation}
P(v) = \frac{1}{\sqrt{2 \pi}{} \sigma_v} \exp\left[- \frac{(v-\langle v \rangle)^2}{2 \sigma_v^2}\right]
\label{eqn:normaldist}
\end{equation}
for a certain range of $\sigma_v$, and the Talbot-lau setups analyzed are at resonance for the mean speed $\langle v \rangle$. 

We have chosen realistic parameters for an experiment with positronium (Ps) atoms subjected to the gravitational acceleration $a=g=9.81 \unit{m\, s^{-2}}$, over a distance $L^{(\rm{TOT})}= 1 \unit{m}$. See Appendix \ref{app:C} for a detailed discussion on the methods and the motivations behind this choice. We also considered the dependence on the open fraction $f$ of the gratings, since the form of the Talbot coefficients indicates that not only the visibility generally depends on $f$, but also that the behaviour of the asymmetric and symmetric setups can be very different for certain values of $f$. In particular, Fig.~\ref{fig:talbotcoeffplots} suggests that  at $f \approx 50 \, \%$ the asymmetric setups could provide an advantage in the visibility. This property is confirmed by our simulations and is physically relevant: in applications where the beam intensity is low (e.g. the inertial sensing of antimatter beams), it is most desirable to employ large open fractions $(f > 30 \, \%)$, in order to maximise the flux. 

\begin{figure}
\centering
    \includegraphics[width=0.37\textwidth]{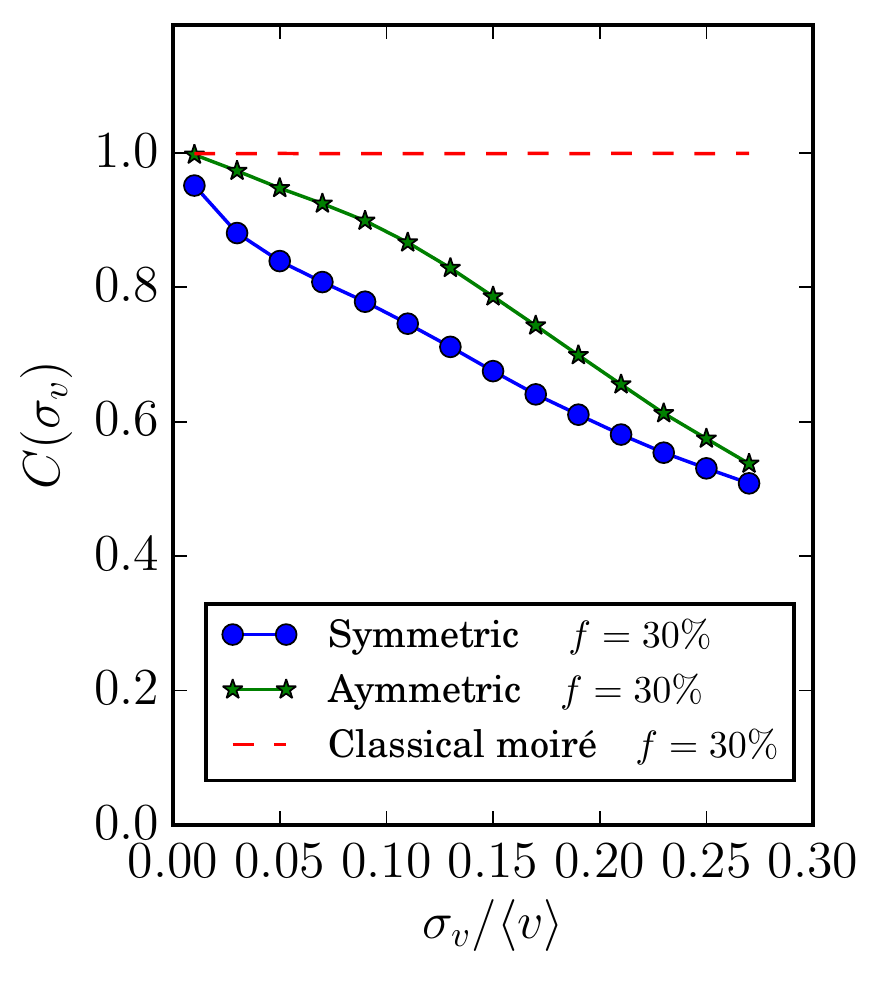}
    \vspace{-0.4cm}
    \caption{Visibility, defined in equation \eqref{eqn:visibilitydefinition}, as a function of $\sigma_v / \langle v \rangle$. The symmetric setup has $d_1=d_2=476 \unit{\mu m}$, $f=0.3$, $L=0.5 \unit{m}$, and $a=9.81 \unit{m s^{-2}}$, whereas the asymmetric setup is defined by $d_1 = 1.5 d_2 = 476 \unit{\mu m}$, $\eta=2$ and $L=0.33 \unit{m}$. These parameters are resonant for a $v=800 \unit{m s^{-1}}$ Positronium atom (see Appendix \ref{app:C}). For comparison we also include a symmetric moiré setup of the same length with $d_1 = d_2 = 800 \unit{\mu m}$. We note that in equation \eqref{eqn:finalresult} only the dimensionless ratios of the grating periods and the parameter $L_T/L \propto v$ appears, so the results on the visibility are general and do not depend on the chosen parameters. }
    \label{fig:visib_nodecay}
\end{figure}

\begin{figure}
\centering
    \includegraphics[width=0.4\textwidth]{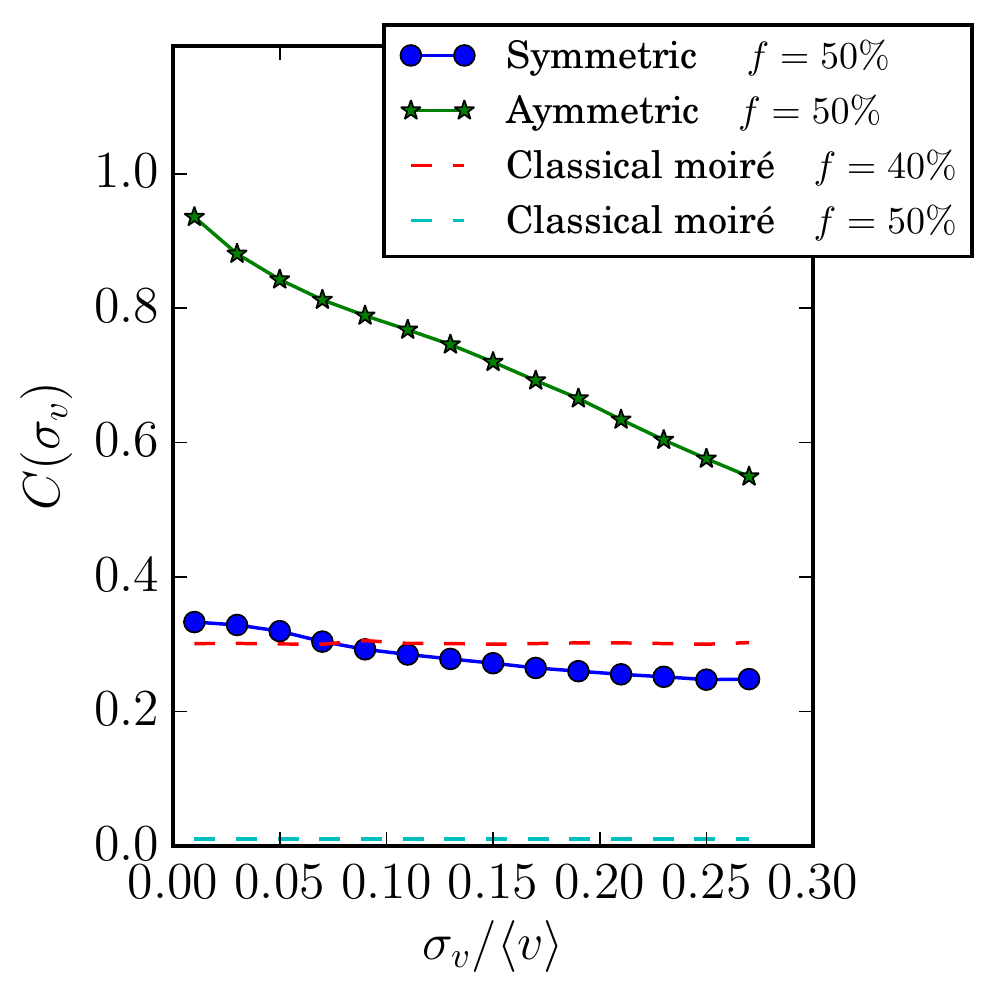}
    \vspace{-0.4cm}
    \caption{Visibility in the same conditions of Fig.~\ref{fig:visib_nodecay}, with an increased open fraction $f=50 \, \%$ and the period of the symmetric setup set to $d_1 = d_1 = 413 \unit{\mu m}$ (see text for details). The dashed lines show the visibility of classical moiré setups for three values of the open fraction: $f=50, 40 \,  \mbox{and} \, 30 \, \%$. Most importantly, we observe that the asymmetric Talbot-Lau setup provides considerably higher contrast than the symmetric setup, for which the value $f=50 \%$ is particularly critical, in analogy with the classical deflectometer \cite{moiree} (see also Fig.~\ref{fig:talbotcoeffplots}). }
    \label{fig:visib_nodecay_50}
\end{figure}

We calculated the contrast $C$ of the intensity patterns via Eq.\eqref{eqn:visibilitydefinition} and the result is shown in Figures \ref{fig:visib_nodecay} and \ref{fig:visib_nodecay_50}, alongside the visibility of the relevant moiré setups for comparison. 

In Fig.~\ref{fig:visib_nodecay} we have set $f=0.3 = 30 \, \%$, and we can see the asymmetric ($\eta=2$) configuration provides a higher visibility than the symmetric setup of the same length. This is a consequence of the fact that it is based on a lower order resonance ($q=1$). In the highly monochromatic case ($\sigma_v/ \langle v \rangle = 1 \% $) both setups match the classical visibility of the moiré setup, which is close to unity at this open fraction. As anticipated, there is no dependence on the speed distribution in the classical case. 

In Fig.~\ref{fig:visib_nodecay_50} we set the open fraction to $f= 50 \%$ and perform the same comparison of Fig.~\ref{fig:visib_nodecay}. However, the period of the symmetric setup has been adjusted to satisfy the appropriate maximum visibility condition at $f=50 \, \%$, that is $L/L_T \approx 1.33$ (see Fig.~\ref{fig:talbotcoeffplots}). In this situation the asymmetric setup provides a more sizeable advantage in visibility, also compared to classical moiré deflectometers with $f=50 \,  \% $ and $ f= 40 \, \%$.
\begin{figure}
\centering
    \includegraphics[width=0.4\textwidth]{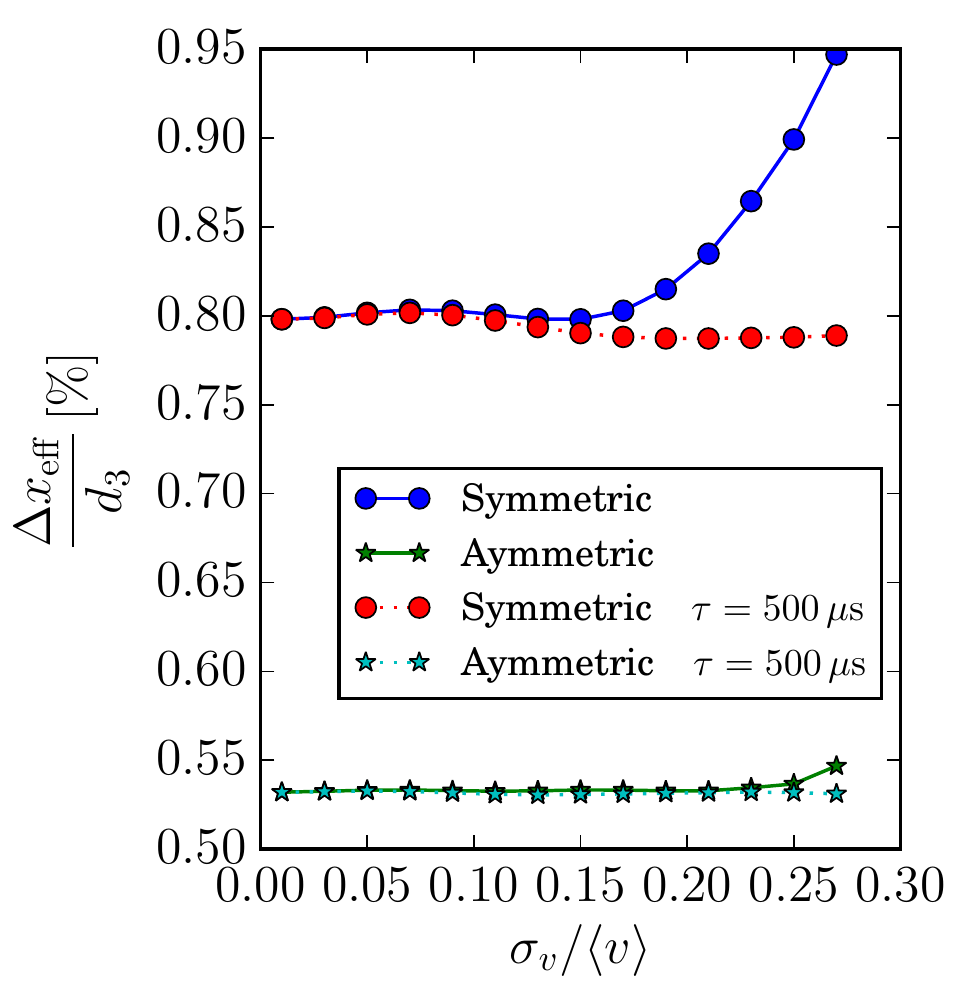}
    \vspace{-0.4cm}
    \caption{Relative displacement $\Delta x_{\rm{eff}}/d_3$ for the same Talbot-Lau configurations described in Fig.~\ref{fig:visib_nodecay}. Both for ideal stable particles, and with a finite lifetime $\tau = 500 \unit{\mu s}$ that alters the effective speed distribution (see the text for details).}
    \label{fig:disp_nodecay}
\end{figure}

While the qualitative features just highlighted are of general validity, we now want to make our description more specific, by considering positronium interferometry and accounting for its finite lifetime $\tau$. The longer lived spin triplet ortho-positronium state has a lifetime $\tau_0 = 142 \unit{ns}$ \cite{pdg} in its ground state, and to devise a Talbot-Lau configuration yielding a measurable displacement under the gravitational acceleration for such a short lived particle is impossible. However the use of excited states of Ps is feasible and has been proposed for this purpose\cite{oberthalerpos,cassidyhogan,giammarchiTalk}. In particular, for high-$n$ Rydberg states \cite{gallagher}, the lifetime scales as $\tau \propto n^2 l^3$ with $n$ and $l$ being the principal and angular quantum numbers respectively, so it is in principle possible to reach lifetimes of the order of $\tau \approx 500 \unit{\mu s}$. We take the finite lifetime into account by assuming that atoms decaying before the detector plane (see Fig.~\ref{fig:asimm2}) are not detected.

We analysed the relative displacement $\Delta x_{\rm{eff}}/d_3$, both in the presence and in the absence of decay, focusing on the $f= 30 \%$ case (the inertial displacement is unaffected by $f$). As seen in Fig.~\ref{fig:disp_nodecay} the symmetric setup provides a larger relative displacement by the factor $\approx 1.3$ predicted by Eq. \eqref{eqn:ratio2} for $\eta=2$. We also observe that, for the symmetric case in particular, there is a sizeable dependence of the effective displacement on $\sigma_v$. This has a physical origin in the fact that, although the maximum variance $\sigma_v$ has been carefully chosen (see appendix \ref{app:C}), as the speed distributions widens the contribution from the slower particles starts to dominate. If one calculates the mean value of the displacement $\Delta x (v) \propto 1/v^2$, equation \eqref{eqn:deltax} for the distribution \eqref{eqn:normaldist}, a parameter that strongly correlates with $\Delta x_{\rm{eff}}$, the same rise appears as a function of $\sigma_v$. The disappearance  of this increase when the particles decay confirms this conclusion: the exponential decay with lifetime $\tau$ produces an effective speed distribution $P_{\rm{eff}}(v)$, different from the one the atoms were initially produced with, namely $P(v)$. This function has the following form:
\begin{equation}
P_{\rm{eff}}(v) \propto P(v) \exp \left( {-\frac{L^{(\rm{TOT})}}{\tau v}} \right),
\label{eqn:effectivespeed}
\end{equation}
it is peaked on a higher speed than $\langle v \rangle$, and it the slower end of the spectrum is suppressed.

\section{Conclusion and final remarks}

\begin{figure}
\centering
    \includegraphics[width=0.4\textwidth]{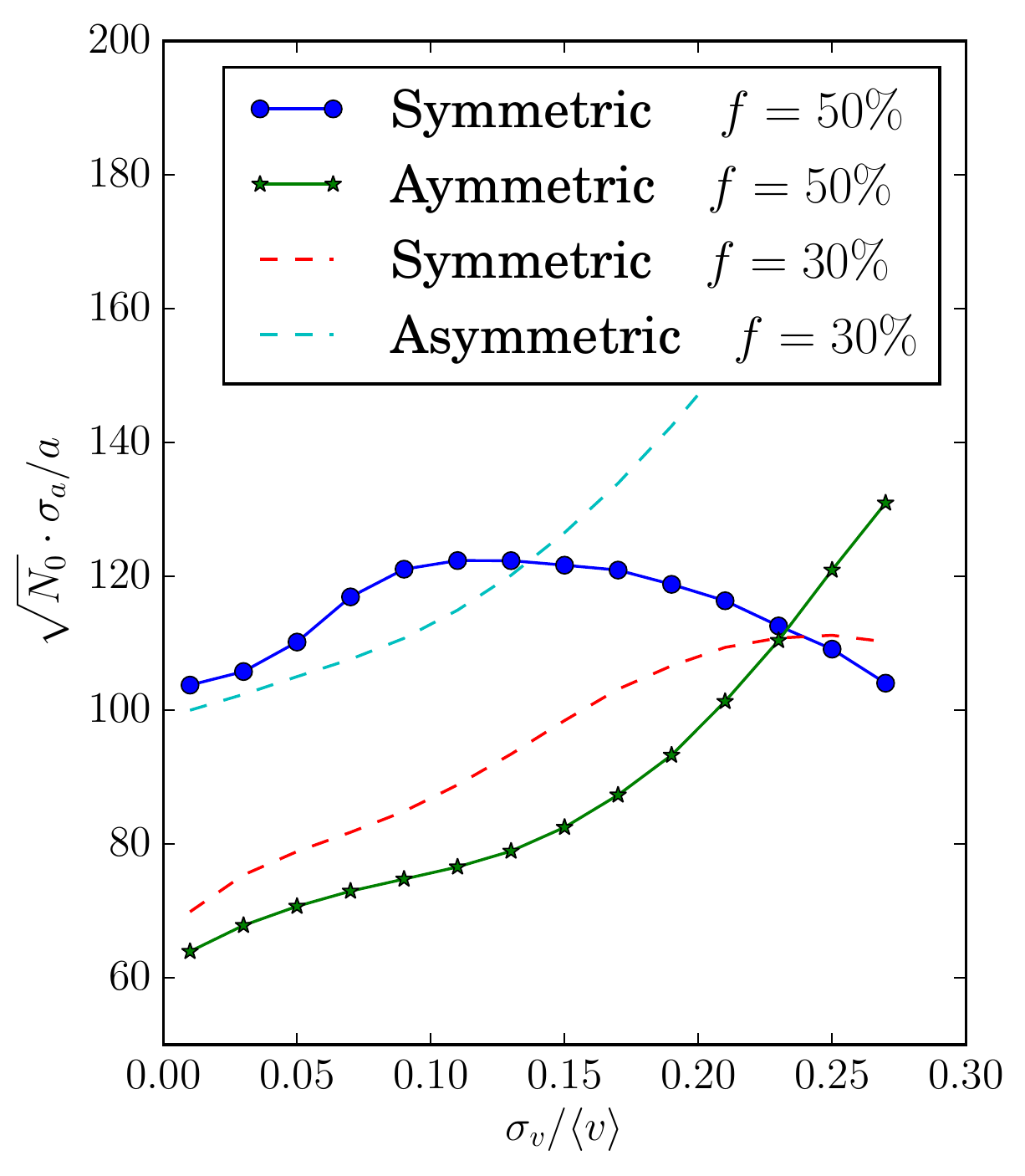}
    \vspace{-0.4cm}
    \caption{  \label{fig:finalresult} Rescaled inertial sensitivity, defined in equation \eqref{eqn:rescaled_inertial}. We compare the four Talbot-Lau configurations of the same total length, $L^{(\rm{TOT})}= 1 \unit{m}$, defined in Figures \ref{fig:visib_nodecay_50} and \ref{fig:visib_nodecay}.}
\end{figure}
To summarise our results, we compare the inertial sensitivity of the four Talbot-Lau configurations we considered, namely the asymmetric $\eta=2$, $L^{(\rm{TOT})}= 1 \unit{m}$ configuration, and the symmetric setup of the same length, at resonance for positronium atoms at $v=800 \unit{m/s}$. We considered in both the cases two values of the open fractions: $f= 30 \%$ and $f=50 \%$ and set the grating periods to achieve the maximum visibility, using the results of section \ref{sec:general_description}. The parameter $f$, in addition to the visibility, also affects the particle flux. In particular it is reasonable to assume that the intensity is proportional to the square of $f$, namely $N=f^2 N_0$. Thus we can define a significant estimator for the inertial sensitivity as:
\begin{equation}
\sqrt{N_0} \left. \frac{\sigma_a}{a} \right|_{N=f^2 N_0},
\label{eqn:rescaled_inertial}
\end{equation}
where $\sigma_a /a$ is defined by equation \eqref{eqn:sensitivityNM}, performing the substitution $N=f^2 N_0$. The impact of the open fraction is thus taken into account. In Fig.~\ref{fig:finalresult}, we plot the function \eqref{eqn:rescaled_inertial} in the absence of decay, that is, for purely Gaussian speed distributions. We can see that throughout most of the $\sigma_v$ range, the best performing configuration is the asymmetric $f=50 \%$ configuration. Moreover, the inertial sensitivity is not the only figure of merit to be considered: the asymmetric setup also provides a larger absolute displacement $\Delta x_{\rm{eff}}$ and interference fringes period $d_3$ by a factor $2 \eta / (\eta +1)$, and $\sqrt{2 \eta}$ respectively (see Eqs.~\eqref{eqn:ratio2} and \eqref{eqn:ratio}). These parameters are always relevant when a finite experimental resolution is taken into account. 

These considerations are of general validity and do not strictly depend on our choice of parameters: due to their more regular behavior (see Fig.~\ref{fig:talbotcoeffplots} and the associated discussion), asymmetric setups can employ higher open fractions, while still matching the visibility of the symmetric setup. This family of resonance conditions were known to exist for the Talbot-Lau interferometer \cite{patorski, beyondeikonal}, but were never studied in detail especially with respect to their inertial sensing capabilities. As a result of our theoretical and numerical analysis, we conclude that the asymmetric Talbot-Lau setups can be very useful, in realistic experimental contexts, to find the optimal compromise between inertial sensitivity, raw statistics, absolute inertial displacement as well as the period of the interference pattern.\\ 

\section*{Acknowledgments}
We would like to thank S. Cialdi and M. Potenza for useful discussions.

\appendix

\section{Derivation of Eq.~\eqref{eqn:finalresult}} \label{app:B}

Here we develop explicitly the steps necessary to evolve the initial state \eqref{eqn:initialwigner} to the observation plane. First of all we note that assuming the second grating has a periodic transmission function $\mathcal{T}_2(x)$, it can be expanded in a Fourier series. Inserting the Fourier decomposition  $\mathcal{T}_2(x) = \sum_n b^{(2)}_n e^{i\frac{2 \pi}{d} n x}$ into equation \eqref{eqn:gratingpropagator} yields the following form for the needed grating transformation:
\begin{align}
K_2(x,x_0; p,p_0) &= \delta (x - x_0) \sum_{k,n}  \e^{i 2 \pi x (n-k) / d_2}\nonumber\\
&\hspace{-1cm}\times b^{(2)}_n \left[ b^{(2)}_k \right]^* \delta\left[p-p_0 - \frac{\pi \hbar}{d}(n+k)\right].
\label{eqn:alternate}
\end{align}
This is the last ingredient needed for the full calculation, which proceeds as in the following scheme (we drop the explicit dependence on $x$ and
$p$):
and
$$
\widetilde{W}_0 \; \xrightarrow[T_1]{G_1 \rightarrow G_2} \;  W_1 \xrightarrow[\mathcal{T}_2(x)]{G_2} 
 \; \widetilde{W}_1 \xrightarrow[T_2]{G_2 \rightarrow \mbox{Screen}}  \; W_2
$$
where we have introduced the Wigner functions immediately before and after the second grating, $W_1(x,p)$ and $\widetilde{W}_1(x,p)$ respectively, and the final state $W_2(x,p)$ from which the intensity distribution at the detection plane is recovered. We remind that we are assuming that the grating slits extend sufficiently in the $z$-direction (the coordinate system is as in Fig.~\ref{fig:asimm2}),so that the problem is effectively one dimensional.

Applying the evolution equation \eqref{eqn:freeevol} and the grating transformation \eqref{eqn:gratingevol} with the form \eqref{eqn:alternate} for the grating convolution function, we get the following expressions:
$$
W_1(x,p)= \widetilde{W}_0 \left( x - \frac{p T_1}{m} + a \frac{T_1^2}{2},p - maT_1 \right)
$$
and
\begin{align*}
\widetilde{W}_1(x,p) &= \int dx_0 dp_0\, K_2(x,p;x_0,p_0) \\
&\hspace{1cm}\times \widetilde{W}_0 \left( x_0 - \frac{p_0 T_1}{m} + a \frac{T_1^2}{2},p_0 - maT_1 \right)
\end{align*}
and finally the state $W_2(x,p)$ after a final free evolution step for a time $T_2$:
\begin{align*}
W_2(x,p) &= \int dx_0 dp_0\, \mathcal{P}\left( \frac{p_0 - maT_1}{p_y} \right) \\
&\hspace{0.5cm}\times  \frac{1}{\mathcal{N} p_y} \left| \mathcal{T}_1(x_0 - \frac{p_0 T_1}{m} + a \frac{T_1^2}{2}) \right| ^2 \\
&\hspace{0.5cm}\times  K_2 \left( x - \frac{p T_2}{m} + a \frac{T_2^2}{2} ,p - maT_2;x_0,p_0 \right).
\end{align*}
We now first apply the following change of variables in the integral (the Jacobian determinant is equal to 1)
\[
\left\{
  \begin{array}{l l}
    p'_0 = p_0 - maT_1\\
    x'_0 =x_0 -p_0 T_1/m + a T_1^2/2
  \end{array} \right.
  \]
then insert the explicit expression \eqref{eqn:alternate} for the convolution factor $K_2$, and integrate over $p$ to get the final position distribution $I(x) = \int W_2(x,p) dp$:
\begin{align}
I(x)&=\sum_{n,k}  \frac{b_n b^*_k}{Gp_y}\int dx'_0 dp'_0 dp \, |\mathcal{T}_1(x'_0)|^2 \mathcal{P} \left( \frac{p'_0}{p_y} \right) \nonumber\\
&\hspace{0.5cm}\times \exp \left\lbrace i\frac{2 \pi (x - p T_2/m + a T_2^2 /2)}{d_2} (n-k) \right\rbrace \nonumber\\
&\hspace{0.5cm}\times \delta \left[ p - ma T_2 - p'_0 - ma T_1 - \frac{\pi \hbar}{d_2}(n+k) \right] \nonumber\\
&\hspace{0.5cm}\times \delta \left( x - \frac{p T_2}{m} + a \frac{T_2^2}{2} - x'_0 - \frac{p'_0 T_1}{m} -a \frac{T_1^2}{2} \right).
\end{align}
After performing the integration over $p$ and $x'_0$, shifting the $k$ summation index as $k' = n-k$, and also introducing the Fourier
series expansion of $|\mathcal{T}_1(x)|^2 = \sum_l A_l \e^{i 2  \pi x l / d_1}$, one obtains:
\begin{align}
I(x) &=\sum_{n,k,l} \int \frac{dp'_0}{\mathcal{N} p_y} \mathcal{P} \left( \frac{p'_0}{p_y} \right) A_l b_n b^*_{n-k} \nonumber \\
&\hspace{0.cm}\times \exp \left\lbrace i \, 2 \pi^2 \frac{(k-2n) \hbar T_2}{m} \left(\frac{ k}{d_2^2} + \frac{l}{d_1 d_2}  \right) \right\rbrace \nonumber  \\
&\hspace{0.cm}\times \exp \left\lbrace i \, 2 \pi \frac{k}{d_2} \left(x - \frac{T_2 p'_0}{m} - a \frac{T_2^2}{2} - a T_1 T_2 \right) \right\rbrace \nonumber  \\
&\hspace{0.cm}\times \exp \left\lbrace i \, 2 \pi \frac{l}{d_1} \left[x - \frac{p'_0}{m} (T_1 + T_2) -\frac{a}{2} (T_1 + T_2)^2 \right] \right\rbrace.
\label{eqn:calcololungo_1}
\end{align}
The Talbot coefficients \eqref{eqn:talbotcoeff} can be recognized into the above integral, with 
$$
\xi =2 \pi \frac{\hbar T_2}{m} \left(\frac{ k}{d_2^2} + \frac{l}{d_1 d_2} \right)
$$
and the scaled Fourier transform of the initial momentum distribution 
$$
\int \frac{dp}{p_y} \mathcal{P} \left( \frac{p}{p_y} \right) \e^{-i p q} \equiv \tilde{\mathcal{P}}(q),
$$
which we can be substituted in \eqref{eqn:calcololungo_1}, to obtain
\begin{align}
I(x)&=\frac{1}{\mathcal{N}} \sum_{k,l} A_l B_k(\xi) \tilde{\mathcal{P}} \left( \frac{2 \pi}{m} \left[ \frac{l}{d_1} (T_1 + T_2) + k \frac{T_2}{d_2} \right] \right) \nonumber \\ &\hspace{0cm}\times \exp \left\lbrace  i \frac{2 \pi}{d_2} \left[ x\left(k+ l \frac{d_2}{d_1}\right) \right] \right\rbrace \nonumber \\
&\hspace{0cm}\times \exp \Bigg\lbrace - i \frac{ 2 \pi}{d_2} \Bigg[  a \Bigg( k T_1 T_2 + k \frac{T_2^2}{2} \nonumber\\
&\hspace{3cm}+ \frac{l d_2}{2 d_1} (T_1 + T_2)^2  \Bigg) \Bigg] \Bigg\rbrace .
\end{align}
To conclude the calculation, we apply a final approximation, namely to assume (as we mentioned in section \ref{sec:general_description}) that the momentum distribution is broad enough that $\tilde{\mathcal{P}}(q) \approx \delta(q)$. Then, by substituting $T_2 = \eta T_1$, which holds in the assumption that the longitudinal motion is unaffected by interference, our Eq.~\eqref{eqn:finalresult} results.

Finally, we remark that the use of Fourier series expansion to define the coefficients \eqref{eqn:talbotcoeff} and $A_l$ is appropriate because the functions $\mathcal{T}_i(x)$ are periodic. To obtain our final result, Eq.~\eqref{eqn:finalresult} it is also assumed that the gratings extend indefinitely in space. This is a reasonable requirement, as long as the number of periods $N$ illuminated by the particle beam is large $N \gtrsim 10^2$. The validity of this approximation can always be checked by calculating the intensity distribution numerically at finite $N$ by means of \eqref{eqn:pointsource} and the Fresnel integral.

\section{Outline of the methods and choice of parameters} \label{app:C}
\begin{figure*}[tb!]
    \includegraphics[width=0.8\textwidth]{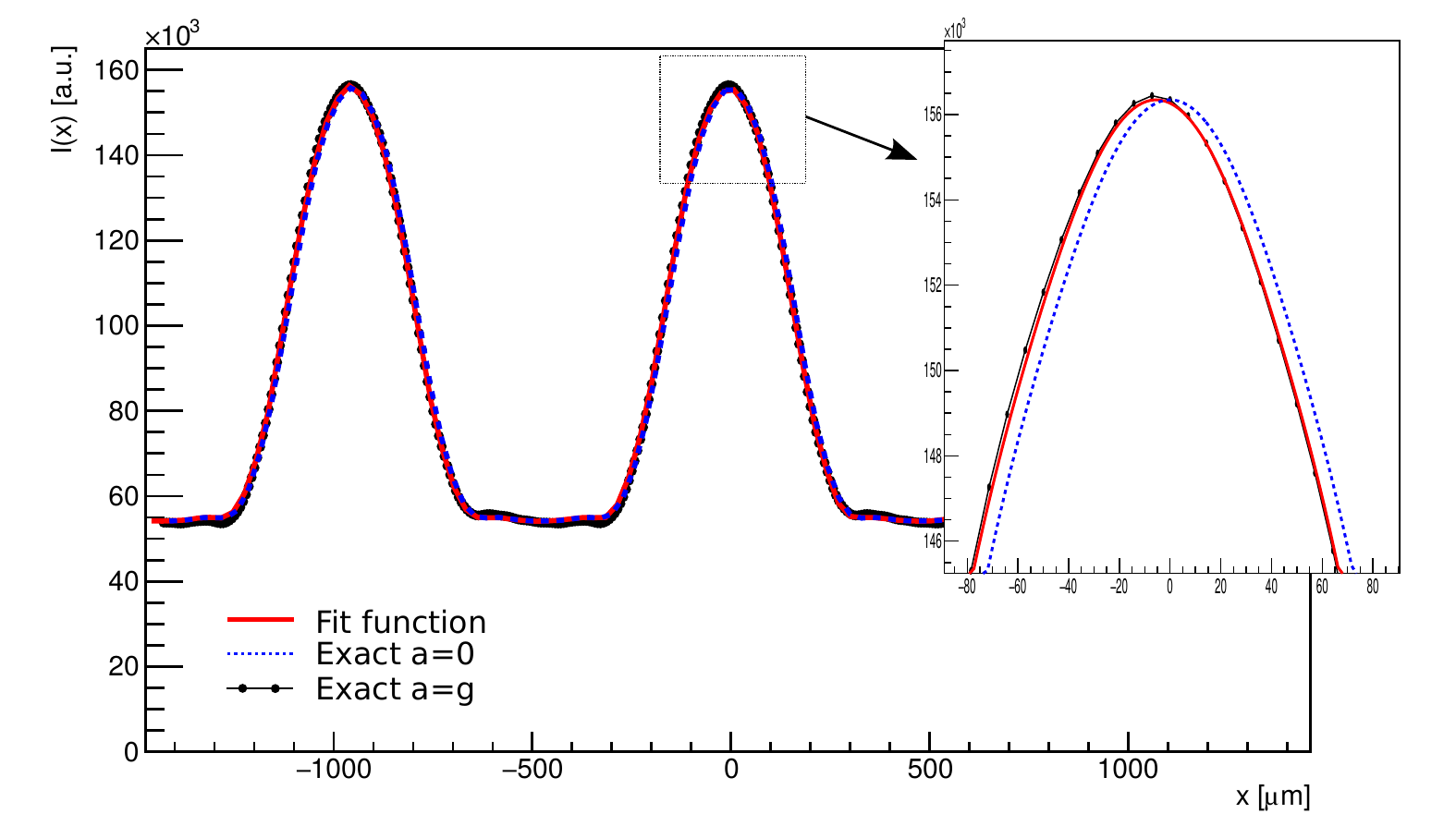}
    \caption{Example result of the fit procedure outlined in Appendix \ref{app:C}. In particular for the case $\sigma_v / \langle v \rangle = 0.3$ and the asymmetric configuration also described in sec \ref{sec:Monte_Carlo}. The box shows a detail of the portion around the peak to highlight the very small displacement of the interference pattern due to the gravitational acceleration with respect to the reference pattern with $a=0$ (with the chosen parameters $\Delta x _{\rm{eff}} = 4 \unit{\mu m}$). It is also evident the good agreement between the fit function, Eq.~\eqref{eqn:intensity_nm_fit}, and the intensity \eqref{eqn:intensity_nm_talbot}.  }
    \label{fig:fit_result}
\end{figure*}
Here we describe in more detail the methods used to obtain the results of section \ref{sec:Monte_Carlo}.
The intensity distribution $I_{\rm{NM}}(x)$ in the presence of the external force and a speed distribution $P(v)$ is evaluated as defined by Eq.~\eqref{eqn:intensity_nm_talbot}. A least squares fit procedure is then performed with the function:
$$
I_{\rm{fit}} (x,x') = \int    I(x - x'|v)  P(v) dv
$$
with the displacement $x'$ being the only free parameter. The displacement $\Delta x _{\rm{eff}}$ is then defined as the best fit value of $x'$, and depends on $P(v)$, hence in our case, $\Delta x _{\rm{eff}} = \Delta x _{\rm{eff}} (\sigma_v)$. By inspecting the results of our numerical analysis (see Fig.~\ref{fig:fit_result}), this is a reliable method to calculate the effective displacement, since the agreement between the fit function and the exact intensity is very good (this justifies the relation between Eq.~\eqref{eqn:intensity_nm_fit} and  Eq.\eqref{eqn:intensity_nm_talbot}), thus being sensitive even to the relative displacements smaller than $ 1 \%$ that we encountered. 
\par
In order for the fit parameter to correspond exactly to the physical displacement we are after, an absolute reference frame has to be established. This is easily done in our computational simulation, by displacing the (monochromatic) intensity function $I(x)$ so that it has an interference peak for the speed $\langle v \rangle$ at $x=0$. For example, for the asymmetric configuration it is necessary to apply a shift of $\eta d_1 /2$. 
\par
While we are interest in a systematic and general comparison of the properties of symmetric and asymmetric Talbot-Lau intertial sensors, we want our choice of parameters to represent a physically relevant case. For this reason, we focus on the possibility to detect the gravitational acceleration $a=9.81 \unit{m s^{-2}}$ of the Positronium (Ps) atom. The possibility to perform quantum interferometry on positronium has been considered in a previous paper \cite{amattint}. Positronium is the bound state of an electron and its antiparticle, having thus a total mass $m_{\rm{Ps}} = 2 m_e$, where $m_e$ is the electron mass. It is an unstable atom with a lifetime $\tau_0 = 142 \unit{ns}$, for the longer lived spin triplet state (ortho-Positronium). 
\par
We focus on a mean speed $ \langle v \rangle = 800 \unit{m/s}$, corresponding to a de Broglie wavelength $\lambda = 454 \unit{nm}$, furthermore we set a total interferometer length $L^{(\rm{TOT})} = 1 \unit{m}$, so that the expected fringe displacement due to the gravitational acceleration on the Earth surface is of the order of a few microns. The chosen velocity distribution is a Gaussian normal \eqref{eqn:normaldist}, whose variance $\sigma_v ^2$ has been chosen so that $\langle v \rangle - 3 \sigma_v >0$, and the Gaussian function is not truncated, to a very good approximation. Therefore in the plots of section \ref{sec:Monte_Carlo}, we are always comparing distribution of the same functional form.   
\par
Our focus was set on two configurations: a symmetric setup \eqref{eqn:resonance_2_symm} with $d_1=d_2 = 476 \unit{\mu m}$, and an asymmetric setup \eqref{eqn:resonance_2_symm} with $d_1 = 1.5 d_2 = 476 \unit{\mu m}$ and $\eta=2$. We chose this low magnification setup because, according to Eq.~\eqref{eqn:relative_asymm} (see also Fig.~\ref{fig:plots_scaling}), the relative displacement of the asymmetric configuration is decreasing with $\eta$. As a matter of fact, we chose a particularly challenging case where the relative gravitational displacement is very small ($\Delta x / d_3 < 1 \%$), due to the small mass of the positronium atom (see Eq.~\eqref{eqn:relative_asymm}). For different experimental conditions (e.g., heavier atoms), the smallness of the relative displacement are not a stringent constraint, and it might be useful to employ also high magnification setups.
\par
To obtain the data in Fig.~\ref{fig:visib_nodecay}, a simple Monte Carlo simulation of a moiré deflectometer has been used. It is based on a direct calculation of the particles trajectories using the classical Eq.~\eqref{eqn:lawsofmotion}, taking into account the transmission properties of the gratings.

%\bibliography{bibliography}

\end{document}